\documentclass[apjl,twocolumn]{openjournal}
\usepackage{xcolor}

\usepackage{textgreek}
\usepackage[utf8]{inputenc}
\usepackage[english]{babel}
\usepackage[flushleft]{threeparttable}

\usepackage{hyperref}
\hypersetup{
    unicode, 
    colorlinks=true,
    linkcolor=linkcolor,
    citecolor=linkcolor,
    filecolor=linkcolor,
    urlcolor=linkcolor,
}
\usepackage{color,colortbl}
\definecolor{linkcolor}{rgb}{0.0,0.3,0.5}
\usepackage{tensind}
\tensordelimiter{?}
\DeclareGraphicsExtensions{.bmp,.png,.jpg,.pdf}
\usepackage{verbatim}
\usepackage[normalem]{ulem}
\usepackage{orcidlink}
\usepackage{soul}

\begin{document}
\title{Distance measurements from the internal dynamics of globular clusters: Application to the Sombrero galaxy (M104)}

\author{Katja Fahrion\orcidlink{0000-0003-2902-8608}$^{1}$}
\author{Michael A. Beasley\orcidlink{0000-0002-4694-2250}$^{2,3,4}$} 
\author{Anastasia Gvozdenko\orcidlink{0000-0002-7619-3131}$^{5}$}
\author{Glenn van de Ven\orcidlink{0000-0003-4546-7731}$^{1}$}
\author{Katherine L. Rhode\orcidlink{0000-0001-8283-4591}$^{6}$}
\author{Ana L. Chies-Santos\orcidlink{0000-0003-3220-0165}$^{7}$}
\author{Anna Ferre-Mateu\orcidlink{0000-0002-6411-220X}$^{2,3,4}$}
\author{Marina Rejkuba\orcidlink{0000-0002-6577-2787}$^{8}$}
\author{Oliver M\"{u}ller\orcidlink{0000-0003-4552-9808}$^{9, 10, 11}$}
\author{Eric Emsellem\orcidlink{0000-0002-6155-7166}$^{8}$}
\email{katja.fahrion@univie.ac.at}
\email{beasley@iac.es}
\affiliation{$^{1}$ Department of Astrophysics, University of Vienna, T\"{u}rkenschanzstra{\ss}e 17, 1180 Wien, Austria}
\affiliation{$^{2}$ Instituto de Astrofísica de Canarias, Calle V\'{i}a L\'{a}ctea, E-38206 La Laguna, Spain.}
\affiliation{$^{3}$ Departamento de Astrof\'{i}sica, Universidad de La Laguna, E-38206 La Laguna, Spain.}
\affiliation{$^{4}$ Centre for Astrophysics and Supercomputing, Swinburne University, John Street, Hawthorn VIC 3122, Australia.}
\affiliation{$^{5}$ Department of Physics, Centre for Extragalactic Astronomy, Durham University, South Road, Durham DH1 3LE, UK.}
\affiliation{$^{6}$ Department of Astronomy, Indiana University, 727 East Third Street, Bloomington, IN 47405, USA.}
\affiliation{$^{7}$ Instituto de F\'{i}sica, Universidade Federal do Rio Grande do Sul (UFRGS), Av. Bento Gon\c{c}alves, 9500, Porto Alegre, RS 91501-970, Brazil}
\affiliation{$^{8}$ European Southern Observatory, Karl-Schwarzschild-Stra{\ss}e 2, 85748 Garching bei M\"unchen, Germany}
\affiliation{$^{9}$ Institute of Physics, Laboratory of Astrophysics, Ecole Polytechnique Fédérale de Lausanne (EPFL), 1290 Sauverny, Switzerland}
\affiliation{$^{10}$ Institute of Astronomy, Madingley Rd, Cambridge CB3 0HA, UK}
\affiliation{$^{11}$ Visiting Fellow, Clare Hall, University of Cambridge, Cambridge, UK}

\begin{abstract}
    Globular clusters (GCs) are dense star clusters found in all massive galaxies. Recent work has established that they follow a tight relation between their internal stellar velocity dispersion $\sigma$ and luminosity, enabling accurate distance measurements. In this work, we aim to apply this GC velocity dispersion (GCVD) distance method to measure the distance to M104 (NGC\,4594, the Sombrero galaxy).
    We have measured internal stellar velocity dispersions for 85 globular clusters (GCs) and one ultra-compact dwarf galaxy around M104 using high-resolution multi-object integrated-light spectroscopy with FLAMES/GIRAFFE on the Very Large Telescope. The measured velocity dispersions range from $\sigma = 4 - 30$ km s$^{-1}$, with a mean uncertainty of $\Delta\sigma = 2.5$ km s$^{-1}$.  For a subset of 77 GCs with $V$-band magnitudes and reliable velocity dispersion measurements above $\sigma > 4$ km s$^{-1}$, we constructed the $M_V$-$\sigma$ relation to measure the distance to M104, finding $D=9.00\pm0.29$ (stat.)~$\pm0.26$ (sys.) Mpc. The GCs follow the Milky Way and M31 $M_V-\sigma$ relation closely, with the exception of the luminous ultra-compact dwarf SUCD1, which is nearly one magnitude brighter than the mean relation. 29 GCs in the sample have sizes determined from {\it Hubble} Space Telescope imaging which allowed us to determine their masses and $V$-band dynamical mass-to-light ratios (M/L$_V$). We find a mean $<M/L_V> = 2.6 M_{\odot}/L_{\odot}$ for the luminous ($M_V < -8$ mag) M104 GCs, which is higher than the Milky Way GCs, but is reminiscent of the brightest GCs in Centaurus\,A. With the exception of SUCD1, the GCs of M104 follow the GCVD relation irrespective of their mass-to-light ratio.
\end{abstract}

% Write your keywords here

\maketitle

\section{Introduction}
\label{sect:intro}
Globular clusters (GCs) are ancient star clusters found in all massive galaxies \citep{Brodie2006, Beasley2020}. With stellar masses $M_\ast = 10^4 - 10^6 M_\sun$ and compact sizes ($r_{\text{h}} = 2 - 10$ pc), they are dense enough to have survived for billions of years. Consequently, as some of the oldest stellar structures in the Universe, GCs have long been employed as ancient fossils of galaxy formation and assembly. 

In the Milky Way, spatially resolved studies have placed constraints on the chemical and kinematic properties of individual stars within GCs. For decades it has been known that GCs follow a tight relation between their magnitudes (or masses) and internal velocity dispersions $\sigma$ \citep{Pryor1993}. Similar to the Faber-Jackson relation for elliptical galaxies \citep{FaberJackson1976}, this relation is a projection of the fundamental plane that arises from the Virial Theorem \citep{Djorgovski1995, Strader2009}: In a self-gravitating system, the total mass scales with $\sigma^{2} r_h$, where $r_h$ refers to the half-light radius. As old stellar systems with a narrow size distribution ($r_h \sim 2 - 5 $pc \citealt{Masters2010}) and a narrow range of mass-to-light ratios (1 - 3; \citealt{McLaughlin2000, BaumgardtHilker2018}), the GC $M_V - \sigma$ relation shows remarkably small scatter. As already discussed by \cite{Paturel1992}, \cite{Djorgovski1995}, and \cite{Strader2009}, this tight relation between GC magnitudes and velocity dispersion can be employed as an accurate distance indicator.

Recently, \cite{Beasley2024} revisited the GC velocity dispersion (GCVD) distance relation by collecting $V$-band magnitudes and velocity dispersions for GCs in the Milky Way and M31 \citep{BaumgardtHilker2018, Strader2011} in order to recalibrate the relation using geometric distances derived from Gaia parallaxes. By then applying the GCVD relation to Centaurus\,A using 57 GCs with velocity dispersion measurements from \cite{Dumont2022}, \cite{Beasley2024} demonstrated that this method can yield distances with uncertainties as low as $3 - 5\%$. Even more recently, the GCVD method was applied to derive a distance to the ultra diffuse galaxy NGC\,1052-DF2, achieving an uncertainty of $\sim 7\%$ based on 5 GCs \citep{Beasley2025}.

As of now, application of the GCVD distance method is limited by the availability of velocity dispersion measurements. Outside the Local Group, dispersion measurements are rare as they rely on moderately high spectral resolution ($R>10,000$) to measure dispersions down to $\sigma \sim 4$ km s$^{-1}$, a typical value for a GC with a mass of $M \sim 10^{5}M_\odot$ (the approximate mean mass of GCs in the Milky Way, \citealt{Baumgardt2020}). 
In an ongoing effort to collect larger samples of GC velocity dispersions, we present here velocity dispersions of 85 GCs and one ultra-compact dwarf galaxy (UCD) around M104 (NGC\,4594, the Sombrero galaxy), based on high-resolution spectroscopy with the FLAMES GIRAFFE multi-object spectrograph at the Very Large Telescope (VLT). In \cite{Fahrion2025b}, we presented an extensive catalogue of GC velocities using the same dataset combined with spectra from the OSIRIS instrument at the Gran Telescopio Canarias (GTC) and analysis of archival VLT MUSE data. In this paper, our focus is on deriving a GCVD distance to M104. Additionally, for a subsample of 29 GCs with available size measurements, we derive dynamical masses and mass-to-light ratios, allowing us to test the effect of sizes and mass-to-light ratios on the $V-\sigma$ plane.

M104 is a particularly interesting target for the GCVD method. As a peculiar galaxy with its prominent, almost edge-on disc, M104 has been studied extensively using space- and ground-based facilities (e.g. \citealt{Emsellem1995, Harris2010, Jardel2011, Gadotti2012, Cohen2020}). Also, its rich GC system has been the focus of many studies (e.g. \citealt{Wakamatsu1977, HarrisHarrisHarris1984, Bridges1992, Bridges1997, Forbes1997, Larsen2002, RhodeZepf2004, Bridges2007, AlvesBrito2011, Dowell2014, Kang2022}). The total number of GCs is estimated to lie between 1500 and 2000 \citep{RhodeZepf2004, Kang2022}, and combining literature samples with newly collected data, we collected line-of-sight velocities for 499 of those in \cite{Fahrion2025b}, corresponding to approximately 25\% of the total GC system. With such a rich GC system, M104 is therefore an ideal target to apply the GCVD method and test it against other distance measurements.

Distance measurements to M104 from the literature range between $\sim$ 7 and 11 Mpc\footnote{From the NASA/IPAC Extragalactic Database. Additional estimates using the Tully-Fisher relation give generally larger distances ($D \sim$ 11 - 22 Mpc; \citealt{Bottinelli1984, Tully1988, Sorce2014}), but it is questionable if this relation can be applied to M104 with its peculiar morphology and gas distribution (see e.g. \citealt{McQuinn2016}).} and encompass a variety of different distance indicators. 
Using the tip of the red giant branch (TRGB) in deep \textit{Hubble} Space Telescope (HST) observations, \cite{McQuinn2016} derived a distance of $D = 9.55 \pm 0.13\,\text{(stat.)} \pm 0.31$ (sys.) Mpc, which is the most recent precise distance measurement to M104. Older distance measurements using surface brightness fluctuations (SBF) based on infrared HST observations, \cite{Jensen2003} reported a distance of 9.08 Mpc, similar to other SBF measurements \citep{Ciardullo1993, Ferrarese2000, Tonry2001} that found distances between 8 and 10 Mpc. Similar values were also reported based on the luminosity functions of GCs (GCLF; \citealt{Bridges1992, Spitler2006}) and planetary nebulae (PNLF; \citealt{Ciardullo1993, Ford1996}).

The aim of this paper is to construct the GCVD relation for M104 GCs for a precise distance measurement. The paper is structured as follows: In Sect. \ref{sect:data}, we describe the FLAMES high-resolution data. Sect. \ref{sect:methods} describes our methods of measuring dispersions, the available photometry and our aperture correction. Sect. \ref{sect:V_sigma_parameters} describes the distribution of M104 GCs on the $V$-$\sigma$ plane. The distance measurement is detailed in Sect. \ref{sect:distance} and compared to literature distances in Sect. \ref{sect:discussion}. We summarise and conclude in Sect. \ref{sect:conclusions}.

\begin{figure*}
    \centering
    \includegraphics[width=0.95\textwidth]{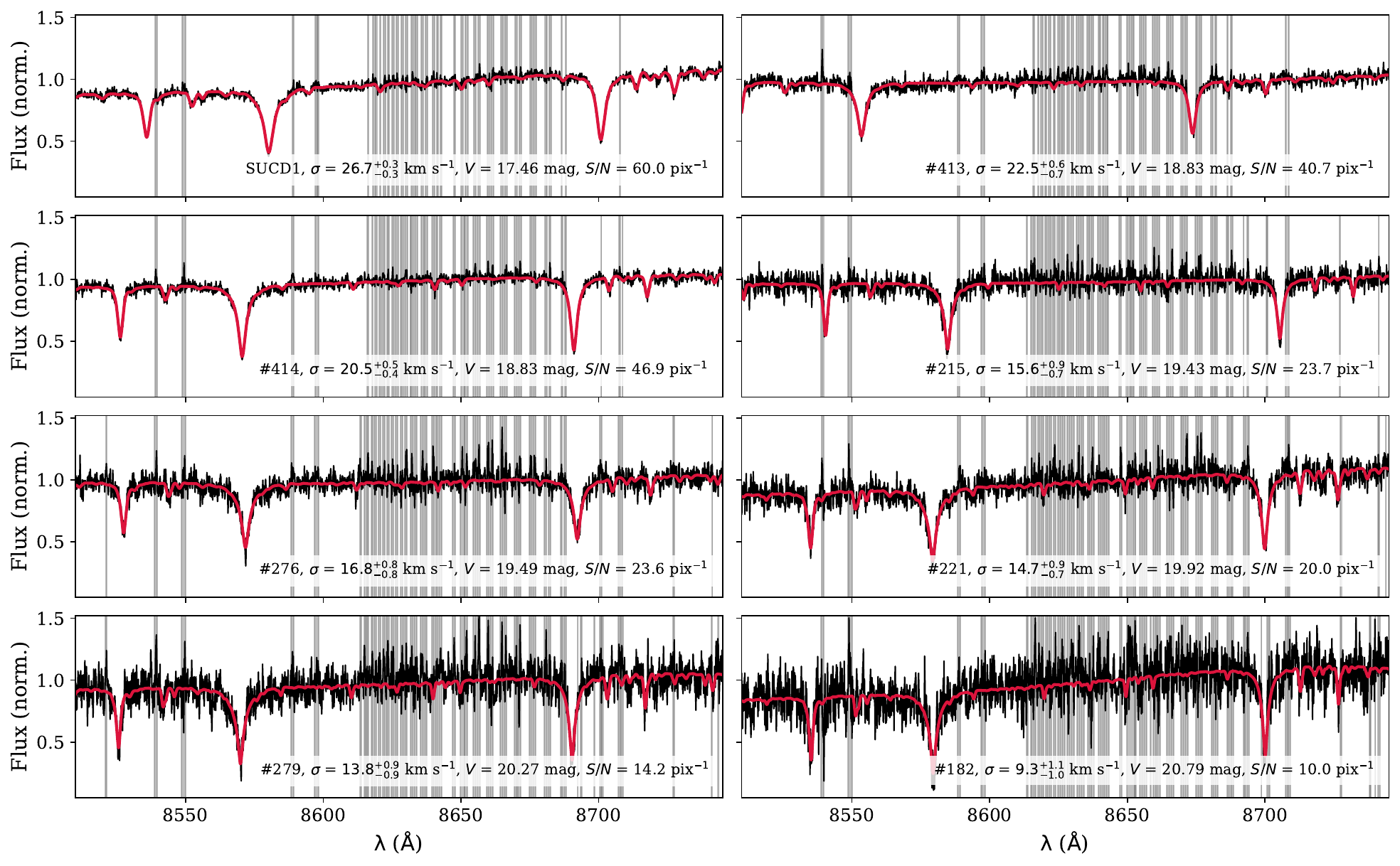}
    \caption{Examples of FLAMES GC spectra. Original spectrum in black, \textsc{pPXF} fit in red. The grey regions indicate the positions of sky lines which were masked out before performing the fits. The GC identifier, the best-fitting velocity dispersion (before aperture correction) and the signal-to-noise ratio are noted in the bottom right corner of each panel.}
    \label{fig:spectra}
\end{figure*}

\section{Data}
\label{sect:data}
The GC system of M104 was observed with FLAMES in January and February 2025 (Programme ID 114.274W, PI: Fahrion). We used the FLAMES instrument with the GIRAFFE spectrograph fed by 135 MEDUSA fibers deployed over the field of view (25\arcmin\,diameter). With two pointings, north and south of the disc, we placed MEDUSA fibers on 221 GC candidates; 93 of them could be identified as GCs bound to M104. SUCD1, a previously identified UCD was also included \citep{Hau2009}. The MEDUSA fibers each span a 1.2\arcsec\,aperture on the sky, corresponding to a physical size of 52 pc at a distance of 9 Mpc (1\arcsec = 44 pc). With a seeing varying from 0.4 to 1.0\arcsec (observation constraints were placed at 1.0\arcsec seeing), the fibers therefore covered most of the GCs' light, but we nonetheless considered aperture corrections as described in Sect. \ref{sect:aperture}.

Full details of the observations and data reduction are given in \cite{Fahrion2025b}. To summarise, we used the FLAMES GIRAFFE HR21 (875nm) grating, which provides coverage of the calcium triplet region from 8483 to 8900 \AA\, with resolution of $R=18,000$ and sampling of 0.05 \AA\,pix$^{-1}$.  Each pointing was covered in ten exposures of 1365 seconds each, distributed over five observing blocks. Targets were selected from the catalogues of \cite{Spitler2006}, \cite{Harris2010} and \cite{Kang2022}.

The data were reduced using standard procedures using the FLAMES GIRAFFE \textsc{esorex} pipeline and custom scripts, with the variation that the sky-subtraction was performed using \textsc{pPXF} \citep{Cappellari2004, Cappellari2017, Cappellari2023} to fit each spectrum with a stellar model from the PHOENIX library \citep{Husser2013} and sky templates as sky input. After sky subtraction, spectra from individual exposures were corrected for barycentric velocities, resampled, and combined using a median weighted by the signal-to-noise (S/N) ratio of each spectrum. 

We also kept a version of spectra with sky lines for masking purposes (see below). Additionally, we used those to test the final spectral resolution of our spectra by fitting prominent sky lines with Gaussian curves.  We found a median full width at half maximum (FWHM) of the sky lines of 0.54 $\pm$ 0.04 \AA. and did not observe any variation in the spectral resolution with wavelength.

\section{Methods}
\label{sect:methods}
In the following, we describe our methods to derive velocity dispersions from the FLAMES spectra. Additionally, we describe the available photometry and how mass-to-light ratios were derived. Sect. \ref{sect:distance} then details the distance measurements.

\subsection{Spectral fitting}
We fit the FLAMES GC spectra with \textsc{pPXF} \citep{Cappellari2004, Cappellari2017, Cappellari2023}. \textsc{pPXF} fits the spectrum with a library of suitable model or empirical spectra and returns the parameters of the line-of-sight velocity distribution (LOSVD) such as radial velocity and velocity dispersion. We chose additive polynomials of degree 4 and multiplicative polynomials of degree 1 to account for variations in the continuum and flux calibrations, but as described in \cite{Beasley2025}, the results are not very sensitive to the choice of degree of the polynomials.
Because sky residuals remained in the reduced data, we carefully masked those lines using a custom mask for every spectrum. We began by normalising the spectrum without sky subtraction in a continuum wavelength range to the continuum level of the sky-subtracted spectrum. Then, we masked out all wavelength pixels in which the normalised sky spectrum exceeds 1.2 times the continuum level. We fitted all spectra in a wavelength range between 8450 and 8850\,\AA\,except for four GCs that had strong sky residuals at redder wavelengths. Here, we restricted the upper wavelength range to 8750 \AA\,due to strong sky residuals at longer wavelengths.

We used the high-resolution stellar population models from \cite{Coelho2014} to fit the FLAMES spectra. In a first step, we selected red (sub) giant stars with effective temperature between 4500 and 5000 K, surface gravity between log(g)= 2.0 and 4.0, and metallicities between $-$1.5 and 0 dex. Then, the selected model spectra were brought to a spectral resolution of 0.54 \AA\,to match the FLAMES spectra. Figure \ref{fig:spectra} shows eight GC spectra and their fits as examples, chosen to cover a range of $V$-band magnitudes and signal-to-noise ratios.

\begin{figure}
    \centering
    \includegraphics[width=0.46\textwidth]{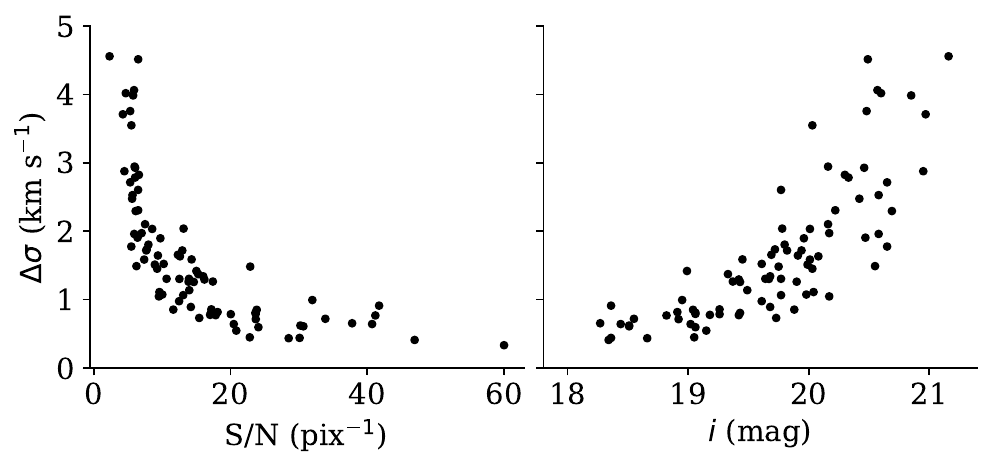}
    \caption{Symmetric velocity dispersion uncertainty as a function of spectral signal-to-noise (left) and $i$-band magnitude (right).}
    \label{fig:snr_vs_sigma}
\end{figure}

As described in \cite{Fahrion2025b}, we performed a Monte-Carlo approach to obtain random uncertainties. For this, 100 representations of the input spectrum were created by drawing values from the residual of the initial fit and adding it onto the initial best fit spectrum. These 100 spectra where then fitted with \textsc{pPXF} for bootstrapping. We used the corresponding histograms of line-of-sight velocities and dispersions to derive the best-fit values and their uncertainties as the 16th, 50th, and 84th percentiles of the distributions. We show the resulting uncertainties on the velocity dispersion as a function of spectral signal-to-noise (S/N) and GC $i$-band magnitude (from \citealt{Kang2022}) in Fig.~\ref{fig:snr_vs_sigma}. From the full sample of 93 GCs, 85 have velocity dispersion uncertainties below 5 km s$^{-1}$ and dispersions $\sigma > 4$ km s$^{-1}$. For the distance measurements, we considered this sample of 85 GCs with reliable velocity dispersion measurements ($\sigma > 4$ km s$^{-1}$, $\Delta\sigma < 5$ km s$^{-1}$).

% \mike{We find good agreement (esp. after aperture correction}

\subsection{Aperture Corrections}
\label{sect:aperture}

The FLAMES velocity dispersions are measured through a finite fibre aperture (1.2\arcsec = 52 pc) and therefore may differ from the true GC velocity dispersions due to incomplete sampling of the GC light. This can occur because light from the outer regions of the cluster may fall outside the fibre, or light from the central regions of the GC is missed due to uncertainties in fibre centering due to either mechanical or astrometrical uncertainties. These  effects were explored in detail in \cite{Beasley2025} where the dual impacts of light falling outside the aperture and imperfect centring on velocity dispersion as a function of seeing, GC size, and distance were simulated.

Briefly, we modelled aperture effects by constructing artificial GCs using modified King profiles \citep{Elson1999}, and then creating velocity dispersion profiles from these models using the Jeans' equation assuming spherical symmetry. After convolution with a range of seeing values, we then compared the fibre-integrated velocity dispersions with the original input values. To explore the effect of imperfect fibre centring, we ran Monte Carlo simulations shifting the artificial GCs within the aperture by $\pm0.1$\arcsec\,and again compared the resulting velocity dispersions with their true values.

For the sizes of M104 GCs ($\sim$3.5 pc; \citealt{Spitler2006, Harris2010}), seeing conditions of our observations (ranging from 0.4 to 1\arcsec) and at the distance of M104 (D$\sim$10 Mpc; \citealt{McQuinn2016}) we find that our observations typically underestimate the true GC dispersions by $\sim 0.4$\%. This effect is dominated by centring uncertainties of the FLAMES fibres. The exception to this is SUCD1, which is sufficiently large to lose some low dispersion light from its outer regions and, in this case our measured dispersion overestimates the true value by $\sim 0.4\%$. While these are small effects, they do have a minor impact on distances measured by the GCVD method. Therefore, where sizes are available we calculate and apply aperture corrections on a cluster by cluster basis. Where no size information is available, we apply a correction based on the average GC size. We also tested if the seeing has a measurable impact on the derived dispersions. For this, we analysed the spectra from individual exposures for particularly bright objects (e.g. SUCD1), but the measured dispersions agree within their uncertainties.

We give the full sample of velocity dispersion measurements in Appendix \ref{ap:ap}. The values range between 4 and 30 km s$^{-1}$ with mean uncertainties of $\sim$2.5 km s$^{-1}$. For SUCD1, we measured an aperture-corrected dispersion of $\sigma=26.6 \pm 0.3$ km s$^{-1}$, in agreement with the measurement by \cite{Hau2009} of $\sigma = 25.0 \pm 5.6$ km s$^{-1}$ based on a lower resolution Keck/DEIMOS spectrum.

\subsection{Available photometry}
\label{sect:available_phot}
\begin{figure}
    \centering
    \includegraphics[width=0.45\textwidth]{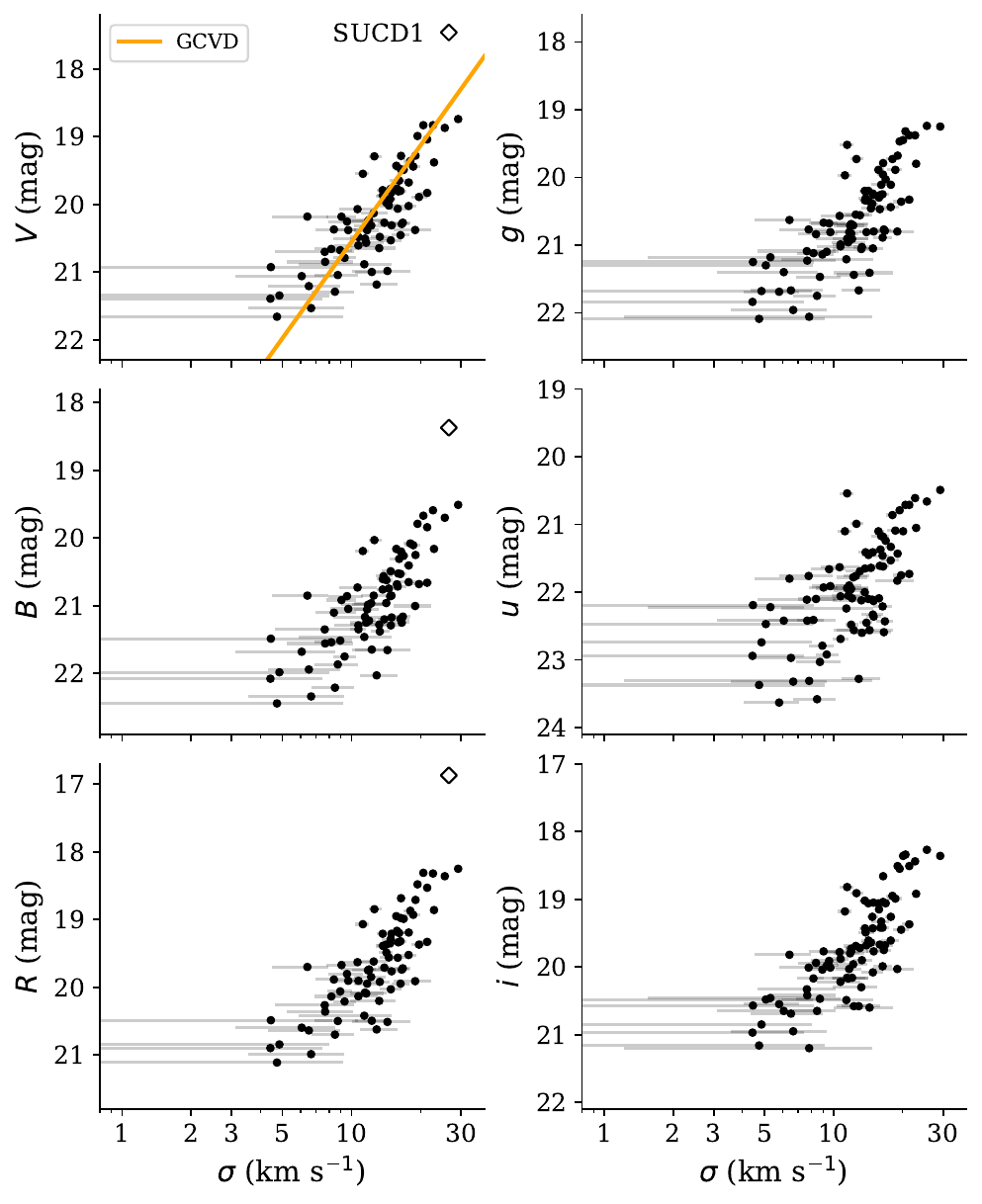}
    \caption{Velocity dispersions versus apparent magnitudes in different filters. The orange line shows the GCVD relation at a distance of 9.00 Mpc. SUCD1 is absent from the right panels as it is not listed in the ground-based catalogue provided in \cite{Kang2022}.}
    \label{fig:mag_vs_sigma}
\end{figure}

\begin{figure*}
        \centering
    \includegraphics[width=0.95\textwidth]{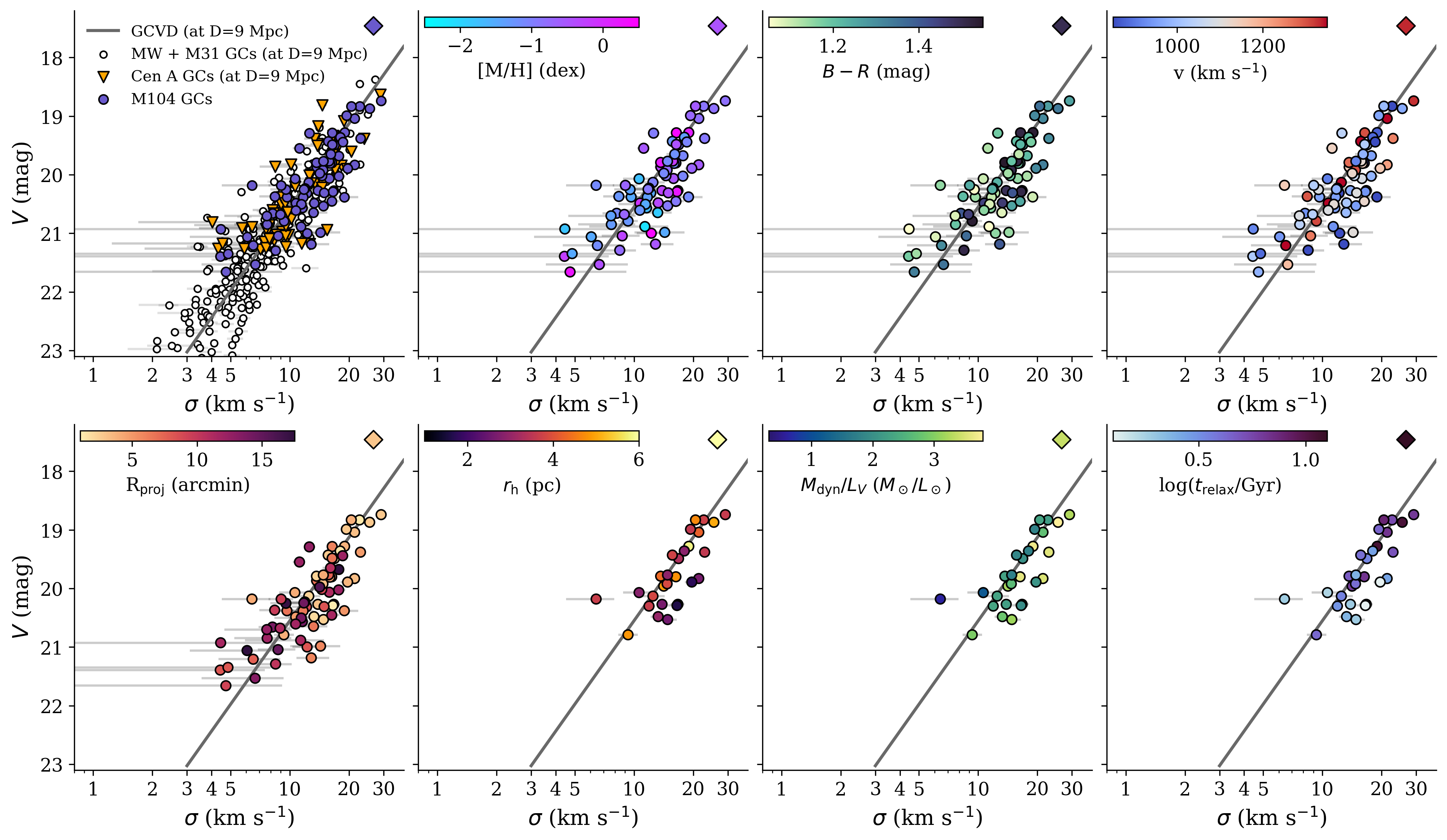}
    \caption{Relationship between apparent magnitude and velocity dispersion, colour-coded with different properties. From top left to bottom right: general distribution compared with MW, M31, and Cen\,A GCs (all placed at a distance of 9.0 Mpc to facilitate comparison with M104), metallicities, $B-R$ colours, line-of-sight velocities, projected distances, half-light radii, mass-to-light ratios, and relaxation times. The solid line is the GCVD relation at a distance of 9.0 Mpc. SUCD1 is marked by the diamond symbol.}
    \label{fig:paramters_on_relation}
\end{figure*}
To directly apply the GCVD approach using the relation calibrated on Local Group GCs, $V$-band magnitudes are needed. For our sample of 85 GCs (not including SUCD1) with velocity dispersion uncertainties $< 5$ km s$^{-1}$ and velocity dispersions $\sigma > 4$ km s$^{-1}$, 29 have high quality $B,V,R$-band magnitudes available from imaging with the Advanced Camera for Surveys (ACS) onboard HST data \citep{Spitler2006, Harris2010}. 

Moreover, 68 GCs have $B,V,R$-band magnitudes available from \cite{RhodeZepf2004} based on imaging with the Mosaic detector on the Kitt Peak 4 metre telescope. The ones with $V$-band magnitudes available from both studies show good agreement within 0.1 mag. In total, $V$-band magnitudes are available for 77 of 85 suitable GCs, in a magnitude range of $V \sim 19 - 22$ mag. For SUCD1, \cite{Hau2009} reported a $V$-band magnitude of 17.46 mag based on the same HST/ACS observations as used by \cite{Spitler2006} and \cite{Harris2010}. In addition, all GCs have $u,g,i$-band magnitudes from \cite{Kang2022} available, based on CFHT/MegaCam imaging. 

Figure \ref{fig:mag_vs_sigma} shows the relation between velocity dispersion and apparent magnitudes in different filters. In all cases, a strong relationship is found due to the underlying relation between mass and dispersion. As the GCVD relation is currently only calibrated on $V$-band magnitudes (shown in the top left panel), we show here the other filters only for visualisation. In Sect. \ref{sect:distance} we also explore distance measurements based conversions from $g$ and $i$ to $V$-band magnitudes.

\section{The Sombrero GCs on the $V$-$\sigma$ plane}
\label{sect:V_sigma_parameters}

Before deriving a GCVD distance, we explore the properties of the GC sample in the $V$-$\sigma$ plane to test underlying dependencies with GC parameters. In Fig.~\ref{fig:paramters_on_relation}, we colour-coded the GCs according to different parameters to test if the scatter correlates with GC parameters. For the full sample, we collected $V$-band magnitudes from the HST study by \cite{Harris2010} and ground-based measurements from \cite{RhodeZepf2004}.

\subsection{Comparison to other GC samples}
The top left panel of Fig.~\ref{fig:paramters_on_relation} compares the $V$ band magnitudes and velocity dispersions of the M104 GCs with that of the Milky Way and M31 (white circles), which were used to calibrate the GCVD relation \citep{Beasley2024}, assuming a distance of 9 Mpc. Additionally, GCs in Centaurus\,A are shown with orange triangles \citep{Dumont2022}. As this panel shows, the M104 GCs cover a very similar range in magnitudes and dispersions as the Cen\,A GCs. The scatter in each direction appears similar to the calibration sample. SUCD1 is about a magnitude brighter than even the brightest GC in the calibration sample or Cen\,A.

\subsection{Metallicities, colours and velocities}
In the second panel on the top left of Fig.~\ref{fig:paramters_on_relation}, the GCs are colour-coded according to their metallicity. The metallicity values are drawn from the full M104 sample described in \cite{Fahrion2025b}. Where available, we used metallicities from optical VLT/MUSE or GTC/OSIRIS spectroscopy, and for the remaining GCs the metallicities were derived directly from full spectrum fitting to the FLAMES spectra. While the GCs with measured dispersions cover a broad range of metallicities, there seems to be no clear correlation with the scatter in the magnitude-dispersion plane. Splitting the MW and M31 GCs into metal-poor and metal-rich GCs, \cite{Beasley2024} found no difference in the GCVD relation's slope or offset, in agreement with our finding for M104.

Similarly, there seems to be no clear correlation between the $B-R$ colour and the scatter on the GCVD relation, as the second panel of Fig.~\ref{fig:paramters_on_relation} shows. Also, the scatter does not seem to correlate with the line-of-sight velocity of the GCs.

\subsection{Projected distances}
The bottom left panel in Fig.~\ref{fig:paramters_on_relation} shows the GCs colour-coded by their projected distance to the centre of M104. The projected distances range from 1 to 20$\arcmin$, and no obvious relation with the scatter in the $V$-$\sigma$ plane is apparent. We note that the brightest GCs in our sample are also relatively close (a few arcminutes) to the centre. This could be a selection effect or could reflect mass segregation of the GC sample since more massive GCs tend to experience stronger dynamical friction effects (e.g. \citealt{Tremaine1976, Turner2012, Lotz2001}). In the Appendix \ref{ap:radial_trends} we show the magnitudes and dispersions as a function of galactocentric distance.

\subsection{Globular cluster sizes}
As the GCVD relation is a consequence of the Virial Theorem, based on the relation between total mass and the combination of velocity dispersion squared and size ($\sigma^2 R_{\text{h}}$), the size influences the position of a source on the relation, with the effect expected to scale linearly. This is seen in the second panel in the bottom row of Fig. \ref{fig:paramters_on_relation}, where smaller-sized GCs lie to the right of the GCVD relation. Also, the observed offset of SUCD1 could be explained by the size difference as it is a particularly large object with a reported half-light radius of $r_{\text{h}} = 14.7 \pm 1.4$ pc \citep{Hau2009}, significantly larger than typical GCs (mean $r_{\text{h}} \sim 2 - 3$ pc, \citealt{Harris2010, Masters2010}).

\begin{figure}
    \centering
    \includegraphics[width=0.45\textwidth]{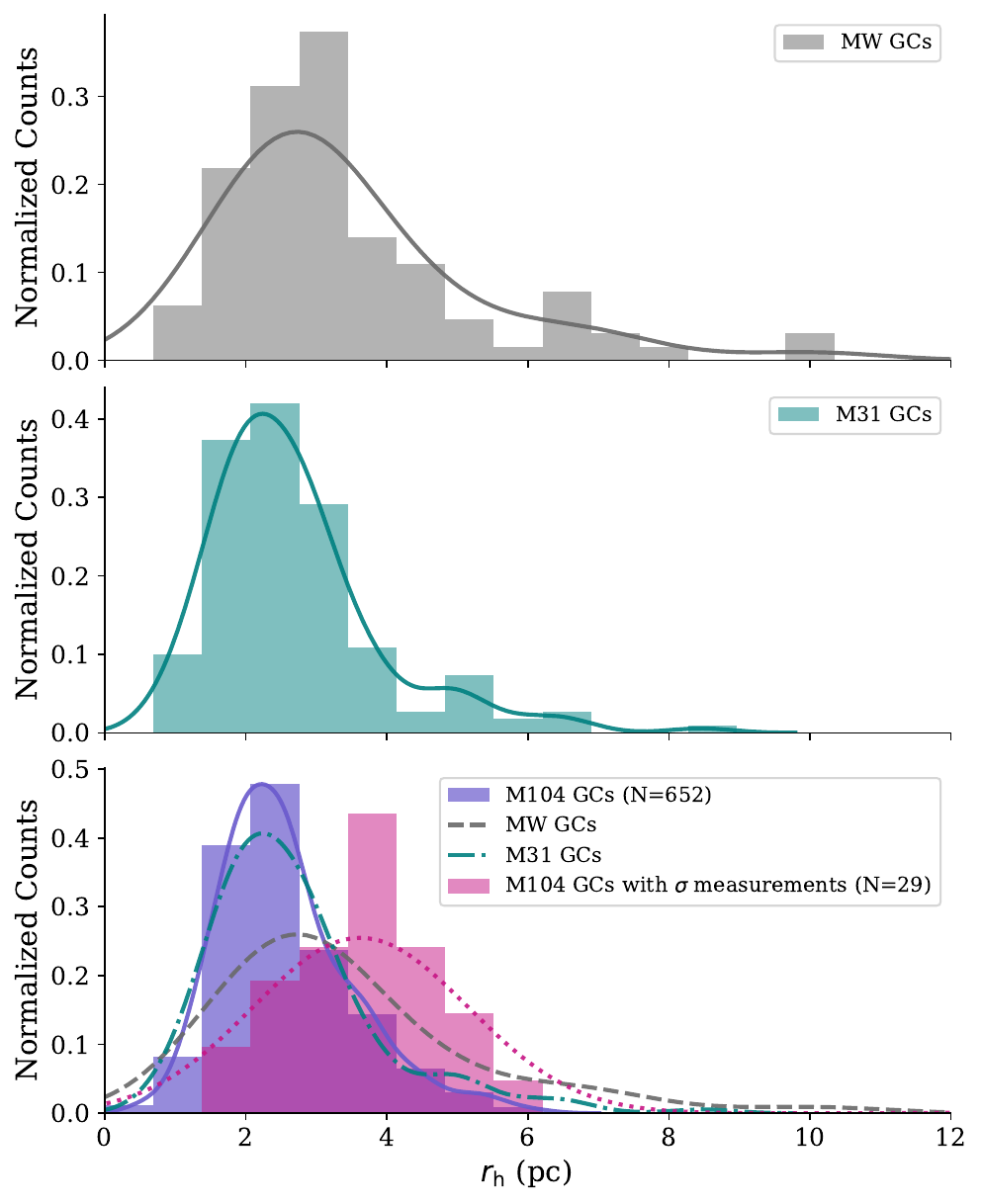}
    \caption{Distribution of GC half-light radii. Top: Milky Way GCs from the catalogue of \cite{BaumgardtHilker2018}. Middle panel: M31 GCs from \cite{Barmby2007}, \cite{Peacock2010}, and \cite{Strader2011}. Bottom: Sizes of M104 GCs from \cite{Harris2010}. The full catalogue of 652 GCs is shown in purple. The pink histogram shows the subsample of 29 GCs for which we have both size and dispersion measurements. The lines in each panel show the kernel densities to better visualise the shape of the distributions.}
    \label{fig:size_distribution}
\end{figure}

Therefore, to test if the GCVD relation is applicable to M104, we compare the sizes of its GCs with the Milky Way (from \citealt{BaumgardtHilker2018}) and M31 GCs (from \citealt{Barmby2007, Peacock2010, Strader2011}) that were used for the original calibration of the relation \citep{Beasley2024}. Figure \ref{fig:size_distribution} shows this comparison, based on the size measurements of 652 M104 GCs from \cite{Harris2010} which were based on HST ACS imaging. We find that the size distribution of M104 GCs aligns well with the sizes of the calibration MW and M31 sample. 

However, we note that the small subsample of 29 GCs for which we have both sizes and dispersion measurements (shown in pink in Fig. \ref{fig:size_distribution}) contains preferentially larger GCs. As these GCs are also among the brighter GCs in our sample (see Fig. \ref{fig:paramters_on_relation}), we believe that this is reflecting a slight brightness-size relation in this small subsample \citep{Harris2010}. The larger sample of GCs with dispersion measurements likely shows a similar size distribution as the calibration sample, but more size measurements would be needed to confirm this.

\subsection{Mass-to-light ratios}
The GCVD distance indicator relies on the assumption that all GCs follow a similar relation between observed light and their mass. To test this assumption, we derived dynamical mass-to-light ratios as described the $V$-band mass-to-light ratios ($M/L_V$) using the aperture-corrected global velocity dispersions ($\sigma_{\rm {gl}}$) and half-light radii ($r_h$) from \cite{Harris2010}. We obtain virial masses for the Sombrero GCs from Eq.~\ref{eq1}~(e.g. \citealt{Larsen2002}; \citealt{Beasley2025}):
\begin{equation}
M_{\rm dyn} = 10 \frac{\sigma_{\rm {gl}}^2 r_{\rm {h}}}{G}
\label{eq1}
\end{equation}

\cite{Beasley2025} showed that Eq.~\ref{eq1} yields good agreement with $M_{\text{dyn}}/L_V$ values determined for Milky Way GCs using spatially resolved stellar dispersion profiles. Therefore, $M_{\text{dyn}}/L_V$ determined via Eq.~\ref{eq1} (e.g. also for the M31 GCs from \citealt{Strader2011} and Cen\,A data from \citealt{Dumont2022}) and $M_{\text{dyn}}/L_V$ measured for the Milky Way GCs should be directly comparable.

\begin{figure}
    \centering
    \includegraphics[width=0.96\linewidth]{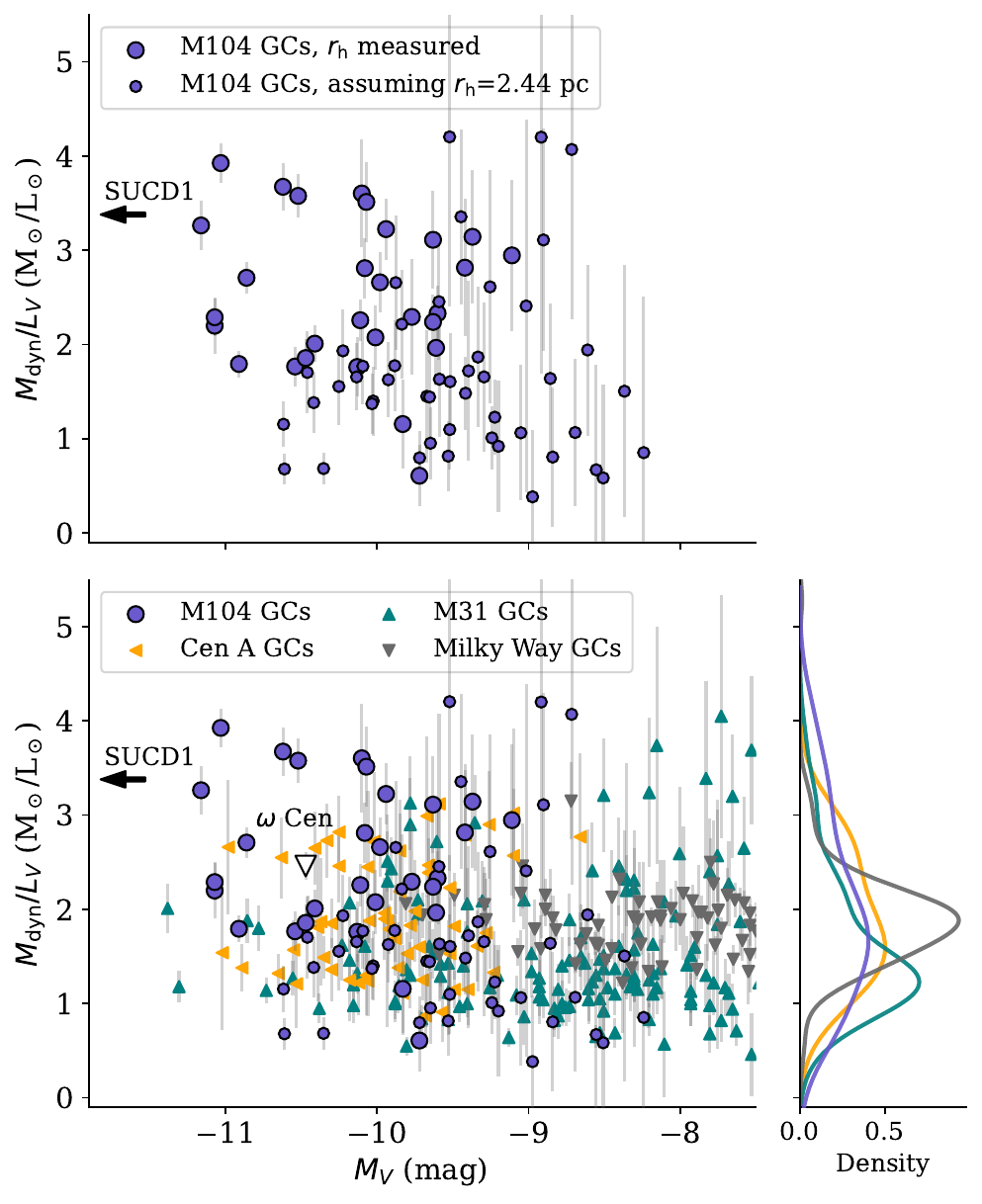}
    \caption{Mass-to-light ratios for the M104 GCs (purple) in comparison with GCs around Centaurus\,A (orange in bottom panel; \citealt{Dumont2022}), M31 (teal; \citealt{Strader2011}), and the Milky Way (grey; \citealt{Baumgardt2020}). $\omega$\,Centauri (NGC\,5139) is marked by the larger grey triangle. SUCD1 is outside the plotted x-range at $M_V = -12.44$ mag, indicated by the arrow. For the M104 GCs, we show the sample of GCs with size measurements as larger circles, and smaller circles refer to GCs with dispersion measurements and $V$-band magnitudes, assuming a half-light radius of $<r_{\text{h}}> = 2.44$ pc, corresponding to the mean of measured sizes. The vertical panel on the right shows the kernel density distributions of $M_{\text{dyn}}/L_V$ for different samples.}
    \label{fig:M_L}
\end{figure}

The derived mass-to-light ratios are shown in Fig.~\ref{fig:M_L} compared to GCs around Centaurus\,A \citep{Dumont2022}, M31 \citep{Strader2011}, and the Milky Way \citep{Baumgardt2020}. We show here both the GCs for which size measurements are available (larger circles) and we also included GCs that lack direct size measurements (smaller circles). For those, we assumed a size of 2.44 pc, corresponding to the median of size measurements as reported in \cite{Harris2010}. These estimated $M/L_{V}$ values show a similar range, with several GCs having similar values as Milky Way GCs. However, we note that the uncertainties increase substantially for the fainter GCs.
For M104 GCs, we find $M_{\text{dyn}}/L_V$ values in a range from $\sim 1 - 4 M_\odot/L_\odot$, with a mean $<M/L_V> = 2.6 M_{\odot}/L_{\odot}$ and a dispersion around the mean of $0.8 M_{\odot}/L_{\odot}$. Many M104 GCs have higher mass-to-light ratios than the MW population, but they are also considerably brighter than most of the Galactic GCs. The brightest M104 GCs are even significantly brighter than $\omega$\,Centauri, the most massive GC of the Milky Way, that also shows an increased mass-to-light ratio.

We find reasonable agreement with GCs in Centaurus\,A ($<M/L_V>_{\text{Cen\,A}} = 2.0 \, M_{\odot}/L_{\odot}$), but a direct comparison is difficult based on the low number of size measurements available for our M104 sample. Similar to the behaviour seen in the Centaurus\,A GCs \citep{Dumont2022}, we also observe a range in mass-to-light ratio that exceeds those of the Milky Way GCs. From their analysis, \cite{Dumont2022} interpreted this scatter with a population of stripped nuclei masquerading as GCs. In M104, we find that SUCD1 indeed shows an elevated mass-to-light ratio, in line with other UCDs (e.g. \citealt{Rejkuba2007, Mieske2008}). Nonetheless, the defining feature that sets it apart from the GC population is its significantly larger luminosity, being about 1.4 magnitudes brighter than the next brightest GC. Based on its magnitude, size, and velocity dispersion, we derive a dynamical mass of $M_{\text{dyn}} = (2.4 \pm 0.2) \times 10^{7} M_{\odot}$, similar to the value reported by \cite{Hau2009}.

The second panel in the bottom row of Fig. \ref{fig:paramters_on_relation} shows the $V$-$\sigma$ plane colour-coded by the $V$-band mass-to-light ratios. GCs with larger mass-to-light ratios appear to lie preferentially to the right of the GCVD relation. Given that the dynamical mass scales with the square of the velocity dispersion, this is perhaps not surprising, but indicates that the GCVD distance method might only by applicable for objects within a narrow range of mass-to-light ratios. Interestingly, SUCD1 is an outlier in this respect as it lies above the relation, but still has a larger mass-to-light ratio than most of the GCs. This is likely a consequence of the  elevated mass-to-light ratio of SUCD1 stemming from its large size, while the higher mass-to-light ratios of the GCs are driven by an increased velocity dispersion.

\subsection{Relaxation times}
The final panel in Fig. \ref{fig:paramters_on_relation} shows the scatter in relation to the half-mass relaxation times $t_{\text{relax}}$, estimated from the sizes and dynamical masses of the GCs. To obtain a measure of $t_{\text{relax}}$, we followed \cite{BinneyTremaine1987}:

\begin{equation}
    t_{\text{relax}} = \frac{0.78\,\text{Gyr}}{\text{ln}(\lambda\,N)} \, \left(\frac{1 M_\sun}{<m>}\right) \, \left(\frac{M}{10^5 M_\sun}\right)^{1/2} \, \left(\frac{r_h}{1 \text{pc}}\right)^{3/2},
\end{equation}
where ln($\lambda N$) is used to describe the Coulomb logarithm, with $N$ as the number of stars within the GC, $<m>$ the average mass of a star, $M$ the mass of the GC, and $r_h$ the half-light radius. We assume $\lambda = 0.1$ \citep{GierszHeggie1994}, $<m> = 0.5 M_\sun$ and \mbox{$N = M/<m>$}. We find relaxation times ranging from $\sim$0.6 to $\sim$ 10 Gyr, while the relaxation time for SUCD1 is estimated at $\sim 88$ Gyr due to its large mass. 

As the relaxation time depends on both the mass and size of the GCs, it is not surprising that the scatter in the GCVD relation shows a similar trend as with the half-light radii: smaller GCs have lower relaxation times and lie to the right of the relation.

\section{Distance to Sombrero}
\label{sect:distance}

\begin{table*}[]
    \centering
    \caption{Distance estimates to Sombrero from different photometric samples}
    \begin{tabular}{l c c} \hline\hline
    Photometry & Distance $\pm$ stat. error & No. GCs \\ 
        & (Mpc) & \\ \hline
          \bf{Combined} $V$-band & \bf{9.00 $\pm$ 0.29} & \bf{77} \\
\cite{Spitler2006}   & 9.33 $\pm$ 0.32 & 29 \\
 \cite{Harris2010}  & 9.00 $\pm$ 0.33 & 29 \\
 \cite{RhodeZepf2004} &  8.89 $\pm$ 0.30 & 68 \\
 $V$ predicted from $g$ \citep{Kang2022} & 9.03 $\pm$ 0.30 & 85\\
 $V$ predicted from $i$ \citep{Kang2022} & 9.14 $\pm$ 0.29 & 85\\ \hline
 
    \end{tabular}
    \label{tab:distances}
\end{table*}

\begin{table}[]
    \centering
    \caption{Sources of systematic uncertainty}
    \begin{tabular}{l l c} \hline\hline
    Source & Uncertainty\\ 
        & (Mpc) & \\ \hline
Aperture/positioning error  & 0.03 ($0.2\%$) \\
Different photometric samples  & 0.16 ($1.7\%$)\\
Different $M_{\text{dyn}}/L_V$ & 0.20 ($2.2\%$) \\ \hline
    \end{tabular}
    \label{tab:systematics}
\end{table}

We obtained a distance to Sombrero using the GCVD approach, which is based on the fact that GCs in the Local Group obey a tight $M_V - \sigma_{\rm gl}$ relation and on the assumption that extragalactic GC systems also follow this relation. See \cite{Beasley2024} and \cite{Beasley2025} for full details of the technique and systematic uncertainties.

\subsection{Distance measurement with the GCVD relation}
We used the same approach as that used in \cite{Beasley2024, Beasley2025} and fit the relation between apparent $V$-band magnitudes and velocity dispersions $\sigma$ with a log-linear relation of the form:
\begin{equation}
    V =  \beta_0 + \beta_1 \text{log}_{10}(\sigma),
    \label{eq:gcvd}
\end{equation}
where $\beta_1$ is the slope of the relation and $\beta_0$ describes the offset. In \cite{Beasley2024}, the relation was calibrated based on Milky Way and M31 GCs, giving values of $\beta_{0,\text{calibrated}} = -4.49 \pm 0.04$ and $\beta_{1,\text{calibrated}} = -4.73 \pm 0.05$. Therefore, when using a fixed slope, the derived offset directly scales with the distance modulus $\mu$:
\begin{equation}
    \mu = \beta_{0,\text{measured}} - \beta_{0,\text{calibrated}}
\end{equation}

As described in \cite{Beasley2025}, we fit the apparent $V$-band magnitudes and velocity dispersions using a Markov Chain Monte Carlo (MCMC) ensemble sampler (\textsc{emcee}; \citealt{emcee}) to determine $\beta_0$ and its uncertainties. We used a Gaussian likelihood function that includes uncertainties in $V$ and $\sigma$. For the offset, we chose a uniform prior in the range of $-10 < \beta_0 < 31$ (corresponding to distances from the Milky Way out to 100 Mpc). 
25 walkers and 10000 chains were used and the corresponding distances describe the median of the posterior distribution with the 16th and 84th percentiles as lower and upper uncertainties. Due to the limited range in magnitudes, we kept the slope fixed to the calibrated relation. In future work, we aim to test the universality of the slope with a larger sample of GC velocity dispersions.

We excluded SUCD1 from all distance measurements due to its peculiar properties. Additionally, it is still debated if UCDs build the high-mass end of the genuine GC population (e.g. \citealt{Bekki2002, Mieske2002}), or are rather the remnant nuclear star clusters (NSCs) of disrupted dwarf galaxies (e.g. \citealt{Bekki2003, PfefferBaumgardt2013}). While they likely are a mixed population (e.g. \citealt{Norris2014, Wang2023}), with its large size and high mass, SUCD1 is likely a NSC-type UCD \citep{Hau2009}. The stripping process \citep{Pfeffer2016} as well as the complex stellar populations found in many NSCs (e.g. \citealt{Kacharov2018, Fahrion2021, Fahrion2024}) might explain its offset from the GCVD relation.

\subsection{Distances from different samples}
\label{subsubsect:samples}

\begin{figure*}
        \centering
    \includegraphics[width=0.93\textwidth]{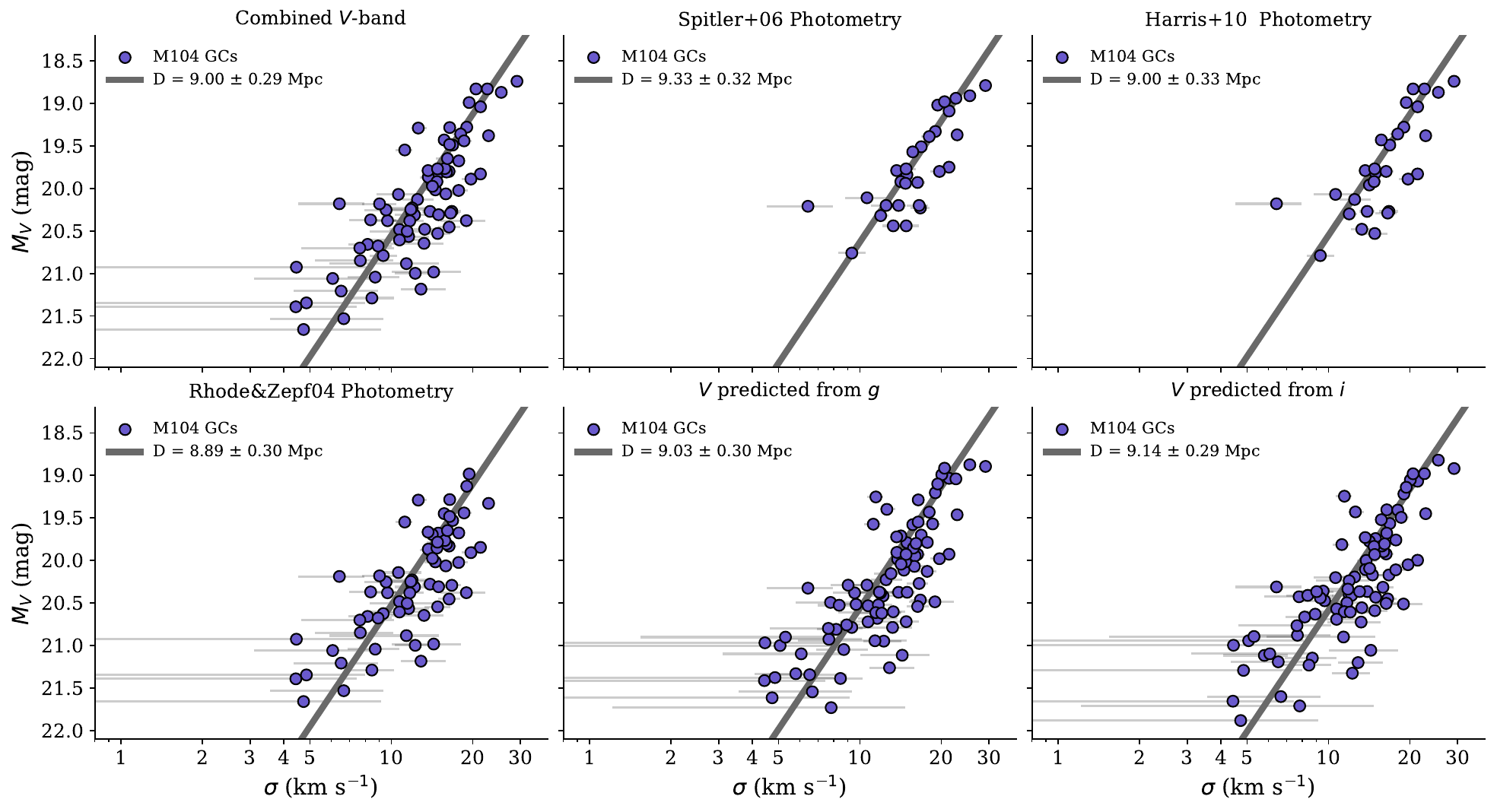}
    \caption{Relationship between absolute $V$-band magnitude and velocity dispersion for the M104 GCs based on different photometric samples. From top left to bottom right: Combined $V$-band photometry, photometry from \cite{Spitler2006} and \cite{Harris2010} based on HST observations, ground-based photometry from \cite{RhodeZepf2004}, and $V$-band magnitudes predicted from the ground-based $g$ and $i$ photometry presented by \cite{Kang2022}. The solid line in each panel indicates the GCVD relation at the best-fitting distance.}
    \label{fig:GCVD_fits}
\end{figure*}

Distance results based on differing photometric samples are presented in Table \ref{tab:distances}.
We can explore the reliability of the GCVD method by dissecting the full GC sample. For example, we can limit the GC sample to only those that have high quality HST-based Johnson $V$-band magnitudes available. Both \cite{Spitler2006} and \cite{Harris2010} provide $V$-band magnitudes based on the same HST/ACS imaging that comprises a 6-pointing mosaic of Sombrero. Overall the photometry from the two studies is in reasonable agreement, although there are differences in the final magnitudes due to different approaches in the aperture correction. \cite{Spitler2006} used a fixed value for all sources, while \cite{Harris2010} used size-dependent aperture corrections. While we have no specific preference for the photometry from either source, the photometry from \cite{Harris2010} might be more reliable as it takes the GC sizes into account and is based on a more detailed consideration of the point spread function that was determined empirically with proximal stars on a cluster-by-cluster basis. There are 29 GCs (excluding SUCD1) in our spectroscopic sample that overlap with both these data sets.  For the photometry from \cite{Spitler2006}, we measure $D=9.33 \pm 0.32$ Mpc, while for the \cite{Harris2010} photometry, we obtain $D=9.00 \pm 0.33$ Mpc. 

We also matched $V$-band photometry from the ground-based study of \cite{RhodeZepf2004} with our sample, yielding 68 GCs. 
\cite{RhodeZepf2004} adopted a single value for the extinction correction of $A_V=0.171$,  while \cite{Spitler2006} adopted $A_V=0.163$.
We therefore adjusted the \cite{RhodeZepf2004} $V$-band magnitudes by -0.008 mag to match the  \cite{Spitler2006} and \cite{Harris2010} photometry.

For this sample we obtain  $D=8.89 \pm 0.30$ Mpc. Therefore, all three samples give similar results, with a standard deviation of only 0.16 Mpc. In view of this, we combined the three samples taking the average of the $V$-band magnitudes in the cases where more than one measurement was available. For this combined sample of 77 GCs, we find $D=9.00 \pm 0.29$ Mpc, which we adopt as our fiducial distance measurement.

Similarly, we can expand the sample by taking the GCs into account that have no $V$-band magnitudes available, but $u,g,i$ photometry from \cite{Kang2022}. To convert their $g$- and $i$-band magnitudes to $V$-band, we made use of the EMILES SSP models \citep{Vazdekis2010, Vazdekis2016}. Applying the appropriate filter transmission curves for CFHT/MegaCam and Johnson $V$ using \textsc{synphot}\footnote{\url{https://synphot.readthedocs.io/en/latest/}} \citep{synphot}, we derived predictions for $V-i$ colours depending on the metallicity of the SSP models, assuming an age of 13 Gyr representative for typical GC systems \citep{Beasley2008, Caldwell2011}. Then, we used these predictions to convert the $g$- and $i$-band magnitudes to $V$-band, based on the measured metallicities \citep{Fahrion2025b}. Deriving the GCVD distance with the $V$-band magnitudes predicted from $g$-band magnitudes yields a distance of $D = 9.03 \pm 0.30$ Mpc, while applying this transformation from $i$- to $V$-band gives a distance of $D = 9.14 \pm 0.29$ Mpc, verifying the validity of the colour conversion.

\subsection{Systematics}
% \kf{[what about the slope? Is that systematic?]}
% \mike{Yes the slope is also a systematic, but it's hard to test it if we don't go faint enough in the sample. We can explore this more fully with larger samples and perhaps by combining sculptor + sombrero for example. For a later paper our perhaps our \"uber paper. For now our "working assumption" is that it doesn't vary much. }
We explored the impact of systematic effects on our distances by considering three key sources of systematic uncertainty: the impact of aperture corrections and fibre centring errors, the effect of using photometry from different sources and samples, and the impact of varying $M_{\text{dyn}}/L_V$. A summary is found in Table \ref{tab:systematics}.

The effect of aperture and fibre centring on the velocity dispersion measurement is discussed in Sect.~\ref{sect:aperture}. These corrections translate to an uncertainty on our distances of 0.03 Mpc (0.2\%) (i.e., this is the difference in distance we obtain with GCVD between the measured velocity dispersions and the aperture-corrected dispersions). Similarly, 
%\kf{[Can you add a few more words about this? How do you get this numner?]}
the impact of different photometry and photometric samples is discussed in Sect.~\ref{subsubsect:samples}, from which we obtain a standard deviation on the three different distances of 0.16 Mpc (1.7\% distance uncertainty). 

To investigate the impact of varying $M_{\text{dyn}}/L_V$ on GCVD, we used the sizes from \cite{Harris2010} where available. To also include the remaining sample, we assumed for the other GCs a mean size of 2.44 pc, the mean of the size measurements. We then performed bootstrap resampling with replacement, where we only selected GCs with $M/L_V<3.0$ to match the Milky Way $M_{\text{dyn}}/L_V$ and M31 distributions. In all cases the GCVD slope was fixed to the fiducial value, and we just consider the impact on the offset $\beta_0$. As a result of this exercise, we find that the offset can shift by up to 0.2 Mpc (2.2\%).

Combining the above mentioned sources of systematics corresponds to a total systematic uncertainty of  $\pm0.26$ Mpc, or 2.9\%.

We also explored the validity of our assumption of using a fixed slope for the GCVD relation obtained from the Milky Way and M31 data. We ran the MCMC fitting on the Sombrero data leaving both the zeropoint and the slope in Eq.~\ref{eq:gcvd} free. As in \cite{Beasley2024} we chose uniform priors in the range $-10<\beta_0<31$ (zeropoint), and $-10<\beta_1<0$ (slope). 
For the full sample, we found $\beta_1 = -5.15\pm0.25$. This would translate to a distance, D = $11.7\pm2.2$ Mpc. This slope is  steeper than that of the fiducial Milky Way + M31 sample ($\beta_1=-4.73\pm0.05$) at the $\sim2\sigma$ level. If we only select Sombrero GCs in the same magnitude range as the fiducial relation (i.e., $M_V > -11$ mag) we find $\beta_1 = -4.98\pm0.23$ and obtain D = $10.5\pm1.6$ Mpc which is consistent with our distance measurement with fixed slope. To conclude, there is some evidence that the $M_V-\sigma$ slope for the brighter Sombrero GCs might be steeper than in the Milky Way and M31 sample, but for the bulk of the sample the assumption of a fixed slope based on fiducial relation is likely reasonable and larger samples across broad magnitude ranges are required to test the effects of the slope properly. 

\begin{figure*}
    \includegraphics[width=0.95\textwidth]{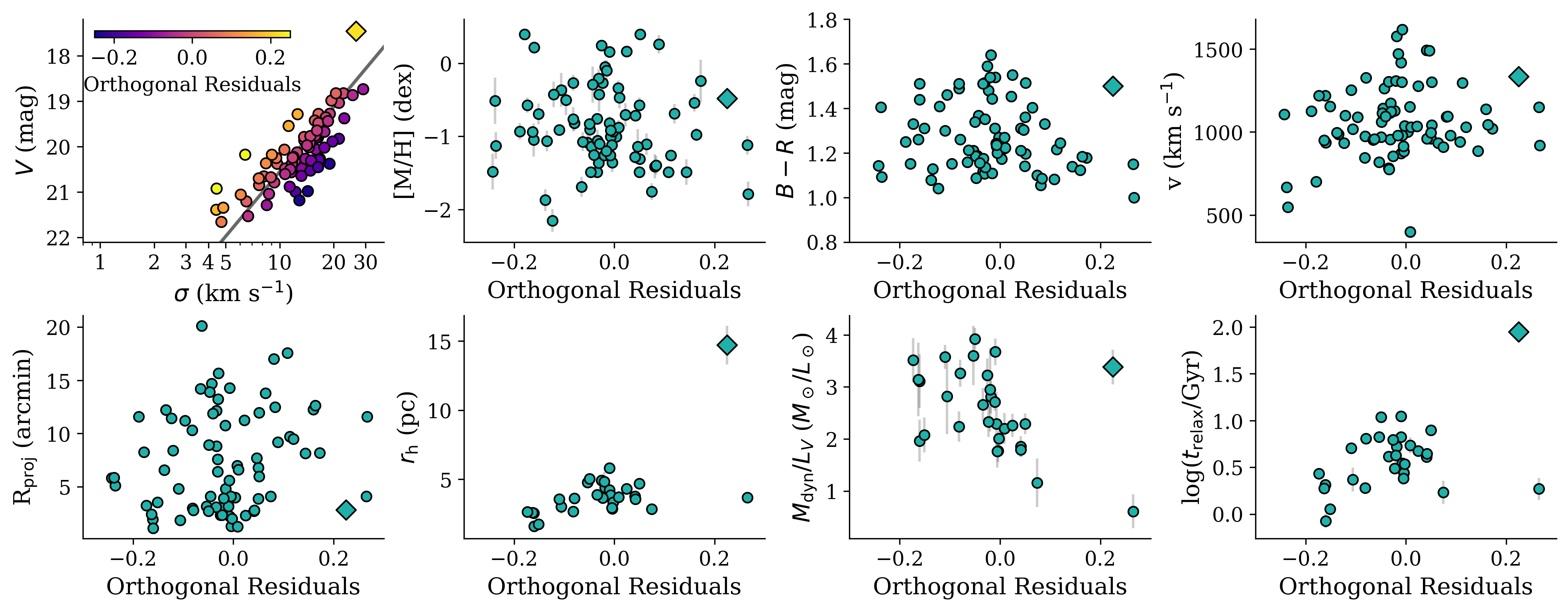}
    \caption{Orthogonal residuals as function of different parameters. The top left panel shows the $V$-$\sigma$ plane with the GCs colour-coded by their orthogonal residuals, which describe the distance to the bestfit GCVD line. The remaining panels show these residuals against various parameters to test for underlying dependences.}
    \label{fig:orthogonal_residuals}
\end{figure*}
\subsection{Dependence on GC parameters}
After fitting the GCVD relation to the M104 GCs, we find a scatter of the residuals of $\sim$ 6\%. This is very similar to the scatter in the GC samples of the Milky Way, M31, and Centaurus\,A (see Fig.~\ref{fig:GCVD_fits}) and suggests that this scatter is intrinsic to such GC samples. In Fig.~\ref{fig:orthogonal_residuals}, we show the orthogonal residuals from the GCVD fit as a function of different GC parameters (similar to Fig. \ref{fig:paramters_on_relation}. The orthogonal residuals refer to distance in $V$ and $\sigma$ to be best-fit line, normalised to the best-fit line.

As can be seen in this figure, the residuals show no dependence on most of the GC parameters such as metallicity, colour, line-of-sight velocity, projected radius or mass-to-light ratio. However, as also noted in Fig. \ref{fig:paramters_on_relation}, there is a weak relation with the size and consequently the relaxation time.

\section{Comparison to literature distances}
\label{sect:discussion}
As a well studied galaxy, multiple distance measurements to M104 exist with varying accuracy. As described in Sect. \ref{sect:intro} and discussed by \cite{McQuinn2016}, distances to M104 vary between $\sim$ 7 and 22 Mpc, with the more accurate measurements placing the galaxy between 8 and 10 Mpc (e.g. \citealt{Jensen2003, Spitler2006, McQuinn2016}). \cite{McQuinn2016} note that distance measurements for M104 based on the Tully-Fisher relation generally find larger distances, between $\sim$ 11 -- 22 Mpc \citep{Bottinelli1984, Tully1988, Sorce2014}. This may be due to the peculiar morphology and spatially restricted gas distribution of M104 that make the morphological classification of this galaxy somewhat uncertain, and set it apart from more typical elliptical and spiral galaxies.

\begin{figure}
    \centering
    \includegraphics[width=0.98\linewidth]{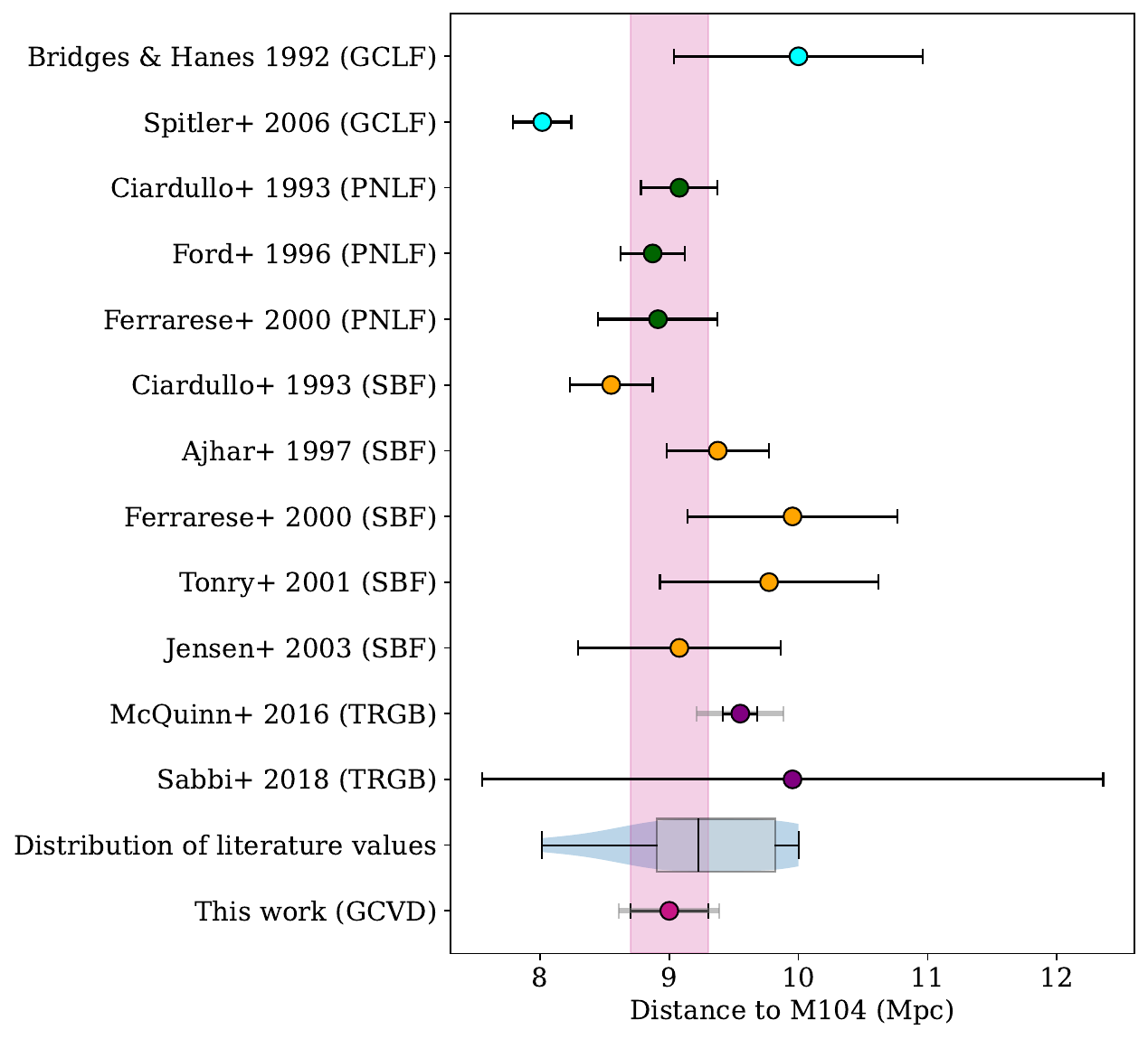}
    \caption{Distance measurements from the literature in comparison to the distance obtained with the GCVD method (pink point). Literature values are grouped according to the employed method. Distances from the GC luminosity function are shown in cyan \citep{Bridges1992, Spitler2006}, green circles refer to measurements from the planetary nebulae luminosity function \citep{Ciardullo1993, Ford1996, Ferrarese2000}, SBF distances are shown in orange \citep{Ciardullo1993, Ajhar1997, Ferrarese2000, Tonry2001, Jensen2003}, and TRGB distances from in purple \citep{McQuinn2016, Sabbi2018}. Black errorbars show 1$\sigma$ statistical uncertainties and the light grey errorbars shows the systematic uncertainties for the TRGB distance from \cite{McQuinn2016} and the GCVD distance, added in quadrature to the statistical uncertainties. The box plot shows the distribution of literature distances, with the vertical black line indicating the median. Distance estimates based on the Tully-Fisher relation were excluded.}
    \label{fig:distances_literature}
\end{figure}

In Fig.~\ref{fig:distances_literature}, we show distances from the literature in comparison with our fiducial GCVD distance of $D = 9.00 \pm 0.29$ (stat.) $\pm$ 0.26 (syst.) Mpc from fitting the combined $V$-band magnitudes (see Sect. \ref{sect:distance}). We show both the statistical uncertainty and the quadrature sum of statistical and systematic uncertainty (in light grey) to visualise a total uncertainty. Distances based on the Tully-Fisher relation were excluded from Fig. \ref{fig:distances_literature} due to the aforementioned reasons. As can be seen from this figure, the GCVD distance compares well against literature measurements from various sources. For example, the median of the shown GCLF, PNLF, SBF, and TRGB distances is at 9.25 Mpc, and calculating the weighted mean taking the statistical uncertainties into account, we find $<D> = 9.09 \pm 0.08$ Mpc, with a dispersion of 0.60 Mpc. 

We find that the GCVD distance agrees very well with the PNLF distances reported by \cite{Ciardullo1993}, \cite{Ford1996}, and \cite{Ferrarese2000} that are all based on the dataset collected by \cite{Ford1996}. While \cite{Ciardullo1993} used a preliminary analysis of the PNLF at that time, \cite{Ferrarese2000} re-computed the PNLF distance after deriving a Cepheid-based calibration for PNLF and other distance indicators. A similar Cepheid-based calibration was used for their SBF distance, based on HST imaging obtained with the wide field planetary camera 2 (WFPC2) described in \cite{Ajhar1997} and later used in \cite{Tonry2001}. \cite{Jensen2003} used new HST Near Infrared Camera and Multi-Object Spectrometer (NICMOS) data for their SBF distance measurement, which is also in agreement with the GCVD distance.

While most distances agree with our GCVD distance within 1$\sigma$ of the combined statistical uncertainties, there are exceptions. For example, the GCLF distance of $D = 8.0 \pm 0.2$ Mpc reported by \cite{Spitler2006} would place the Sombrero galaxy significantly closer than our GCVD distance. As \cite{Spitler2006} discuss, assuming a distance of 9.0 Mpc would make the turnover magnitude of the GCLF about 0.3 mag brighter than the universally adopted value. However, investigation of early-type galaxies in the Virgo and Fornax galaxy clusters have found that massive galaxies often have brighter GCLF turnover magnitudes (by $\sim$ 0.1 - 0.3 mag compared to low-mass galaxies, \citealt{Villegas2010}). The bright turnover magnitude in M104 might therefore be a consequence of M104 being a massive galaxy ($M_\ast \sim 2 \times 10^{11} M_\sun$; \citealt{Jardel2011, Karunakaran2020}).

Comparing the GCVD relation to the TRGB distance of $D = 9.55 \pm 0.13$ Mpc from \cite{McQuinn2016} also shows an offset within 2$\sigma$, mostly due to the small statistical uncertainties of the TRGB measurement. Their precise value is based on dedicated, deep HST/ACS data, but \cite{McQuinn2016} quote a systematic uncertainty of $\pm 0.31$ Mpc stemming from the TRGB calibration \citep{Carretta2000, Rizzi2007}. Under consideration of this systematic uncertainty, the TRGB and GCVD distances agree. \cite{Sabbi2018} also derived a TRGB distance to M104 based on HST Wide Field Camera 3 (WFC3) data and they found a larger distance of $D \sim 10$ Mpc. However, given that their data is shallower, they caution that the location of the TRGB is not clearly detected in their data. 
In the future, additional observations with JWST could be used to measure precise TRGB and SBF distances (e.g. \citealt{Anand2024, Jensen2025}).

\section{Conclusions}
\label{sect:conclusions}

In this paper, we employed the relationship between GC luminosities and their internal velocity dispersions to derive the distance to M104 using the GCVD relation. We summarise our finding as follows: 
\begin{itemize}
    \item Using FLAMES high resolution spectroscopy, we derived velocity dispersions of 93 GCs and one UCD. Of those, 85 GCs have $\sigma > 4$ km s$^{-1}$ and velocity dispersion uncertainties \mbox{$<$ 5 km s$^{-1}$}, which we used to derive the GCVD-based distance.
    \item Using $V$-band magnitudes from space- and ground-based observations of 77 GCs, we applied the GCVD relation with fixed slope and obtained a distance of $D=9.00 \pm 0.29$ (stat.) $\pm$ 0.26 (sys.) Mpc.  We find an intrinsic scatter of $\sim$ 6\%. Testing different photometric samples including $V$-band magnitudes derived from other filters finds consistent distance measurements. 
    \item We find that SUCD1, a known UCD near M104, does not follow the GCVD relation and instead shows an offset. This may be related to its larger size or possibly more complex stellar population than the GCs we studied. With a reported half-light radius of 14 pc, SUCD1 is significantly larger than typical GCs.
    \item For a subsample of 29 GCs with available size measurements, we derived dynamical mass-to-light ratios, finding a range of values from $M_{\text{dyn}}/L_V \sim 1 - 4\, M_{\odot}/L_{\odot}$. Similar values were found for GCs in the halo of Centaurus\,A \citep{Dumont2022}, while GCs in the Milky Way typically have lower $M_{\text{dyn}}/L_V$ values, but also have lower masses than the studied M104 GCs. Besides SUCD1, other GCs with elevated mass-to-light ratios follow the GCVD relation.
    \item Our best-fitting distance of $D = 9.00 \pm 0.29$ Mpc agrees well with previous distance measurements in the literature that typically range from 8 to 10 Mpc. Our GCVD distance places M104 slightly closer than the TRGB distance of $D = 9.55 \pm 0.13\,\text{(stat.)}\,\pm\,0.31\,\text{(sys.)}$ Mpc reported in \cite{McQuinn2016}, which currently has the lowest reported statistical uncertainty but a larger systematic uncertainty. Nonetheless, the two distance measurements agree within 2$\sigma$.
\end{itemize}

With velocity dispersion measurements of 85 GCs, the sample presented here is already the second largest sample of GC velocity dispersion measurements beyond the Local Group after Centaurus\,A. In the future, we plan to collect additional high-resolution spectroscopy of extragalactic GC systems to explore the GCVD method, its applicability and underlying systematics with larger samples.

\section*{Acknowledgments}
We thank the two referees for constructive reports that have helped to improve this manuscript. KF thanks Francisco Aros for helpful discussions. KF acknowledges funding from the European Union’s Horizon 2020 research and innovation programme under the Marie Sk\l{}odowska-Curie grant agreement No 101103830. ACS acknowledges support from FAPERGS (grants 23/2551-0001832-2 and 24/2551-0001548-5), CNPq (grants 314301/2021-6, 312940/2025-4, 445231/2024-6, and 404233/2024-4), and CAPES (grant 88887.004427/2024-00). A.F.M acknowledges support from RYC2021-031099-I of MICIN/AEI/10.13039/501100011033/ UE NextGenerationEU/PRTR. O.M. is grateful to the Swiss National Science Foundation for financial support under the grant number PZ00P2\_202104. KF, MB, AFM, ACS, GvV, AG acknowledge support from PID2024-16088NB-I00.
This work made use of Astropy:\footnote{\url{http://www.astropy.org}} a community-developed core Python package and an ecosystem of tools and resources for astronomy \citep{astropy2013, astropy2018, astropy2022}. Based on observations collected at the European Southern Observatory under ESO programme 114.274W (PI: Fahrion).

\bibliographystyle{mnras}

% You should give the same name for your .bbl as your main .tex
% since it is a requirement for posting on ArXiv.
\bibliography{references}

@INPROCEEDINGS{Elson1999,
       author = {{Elson}, R.~A.~W.},
        title = "{Stellar dynamics in globular clusters.}",
    booktitle = {Globular Clusters},
         year = 1999,
       editor = {{Mart{\'\i}nez Roger}, C. and {Perez Fourn{\'o}n}, I. and {S{\'a}nchez}, F.},
        month = jan,
        pages = {209-248},
       adsurl = {https://ui.adsabs.harvard.edu/abs/1999glcl.conf..209E}
}

@ARTICLE{Dumont2022,
       author = {{Dumont}, Antoine and {Seth}, Anil C. and {Strader}, Jay and {Voggel}, Karina and {Sand}, David J. and {Hughes}, Allison K. and {Caldwell}, Nelson and {Crnojevi{\'c}}, Denija and {Mateo}, Mario and {Bailey}, John I. and {Forbes}, Duncan A.},
        title = "{A Population of Luminous Globular Clusters and Stripped Nuclei with Elevated Mass to Light Ratios around NGC 5128}",
      journal = {\apj},
         year = 2022,
        month = apr,
       volume = {929},
       number = {2},
          eid = {147},
        pages = {147},
          doi = {10.3847/1538-4357/ac551c},
archivePrefix = {arXiv},
       eprint = {2112.04504},
 primaryClass = {astro-ph.GA},
       adsurl = {https://ui.adsabs.harvard.edu/abs/2022ApJ...929..147D}
}

@ARTICLE{Norris2014,
       author = {{Norris}, Mark A. and {Kannappan}, Sheila J. and {Forbes}, Duncan A. and {Romanowsky}, Aaron J. and {Brodie}, Jean P. and {Faifer}, Favio Ra{\'u}l and {Huxor}, Avon and {Maraston}, Claudia and {Moffett}, Amanda J. and {Penny}, Samantha J. and {Pota}, Vincenzo and {Smith-Castelli}, Anal{\'\i}a and {Strader}, Jay and {Bradley}, David and {Eckert}, Kathleen D. and {Fohring}, Dora and {McBride}, JoEllen and {Stark}, David V. and {Vaduvescu}, Ovidiu},
        title = "{The AIMSS Project - I. Bridging the star cluster-galaxy divide$^{★}${\textdagger}{\textdaggerdbl}{\textsection}{\textparagraph}}",
      journal = {\mnras},
         year = 2014,
        month = sep,
       volume = {443},
       number = {2},
        pages = {1151-1172},
          doi = {10.1093/mnras/stu1186},
archivePrefix = {arXiv},
       eprint = {1406.6065},
 primaryClass = {astro-ph.GA},
       adsurl = {https://ui.adsabs.harvard.edu/abs/2014MNRAS.443.1151N}
}

@ARTICLE{Bekki2003,
       author = {{Bekki}, K. and {Couch}, W.~J. and {Drinkwater}, M.~J. and {Shioya}, Y.},
        title = "{Galaxy threshing and the origin of ultra-compact dwarf galaxies in the Fornax cluster}",
      journal = {\mnras},
         year = 2003,
        month = sep,
       volume = {344},
       number = {2},
        pages = {399-411},
          doi = {10.1046/j.1365-8711.2003.06916.x},
archivePrefix = {arXiv},
       eprint = {astro-ph/0308243},
 primaryClass = {astro-ph},
       adsurl = {https://ui.adsabs.harvard.edu/abs/2003MNRAS.344..399B}
}

@ARTICLE{PfefferBaumgardt2013,
       author = {{Pfeffer}, J. and {Baumgardt}, H.},
        title = "{Ultra-compact dwarf galaxy formation by tidal stripping of nucleated dwarf galaxies}",
      journal = {\mnras},
         year = 2013,
        month = aug,
       volume = {433},
       number = {3},
        pages = {1997-2005},
          doi = {10.1093/mnras/stt867},
archivePrefix = {arXiv},
       eprint = {1305.3656},
 primaryClass = {astro-ph.GA},
       adsurl = {https://ui.adsabs.harvard.edu/abs/2013MNRAS.433.1997P}
}

@ARTICLE{Mieske2002,
       author = {{Mieske}, S. and {Hilker}, M. and {Infante}, L.},
        title = "{Ultra compact objects in the Fornax cluster of galaxies: Globular clusters or dwarf galaxies?}",
      journal = {\aap},
         year = 2002,
        month = mar,
       volume = {383},
        pages = {823-837},
          doi = {10.1051/0004-6361:20011833},
archivePrefix = {arXiv},
       eprint = {astro-ph/0201011},
 primaryClass = {astro-ph},
       adsurl = {https://ui.adsabs.harvard.edu/abs/2002A&A...383..823M}
}

@ARTICLE{Bekki2002,
       author = {{Bekki}, K. and {Forbes}, Duncan A. and {Beasley}, M.~A. and {Couch}, W.~J.},
        title = "{Globular cluster formation from gravitational tidal effects of merging and interacting galaxies}",
      journal = {\mnras},
         year = 2002,
        month = oct,
       volume = {335},
       number = {4},
        pages = {1176-1192},
          doi = {10.1046/j.1365-8711.2002.05708.x},
archivePrefix = {arXiv},
       eprint = {astro-ph/0206008},
 primaryClass = {astro-ph},
       adsurl = {https://ui.adsabs.harvard.edu/abs/2002MNRAS.335.1176B}
}

@ARTICLE{Wang2023,
       author = {{Wang}, Kaixiang and {Peng}, Eric W. and {Liu}, Chengze and {Mihos}, J. Christopher and {C{\^o}t{\'e}}, Patrick and {Ferrarese}, Laura and {Taylor}, Matthew A. and {Blakeslee}, John P. and {Cuillandre}, Jean-Charles and {Duc}, Pierre-Alain and {Guhathakurta}, Puragra and {Gwyn}, Stephen and {Ko}, Youkyung and {Lan{\c{c}}on}, Ariane and {Lim}, Sungsoon and {MacArthur}, Lauren A. and {Puzia}, Thomas and {Roediger}, Joel and {Sales}, Laura V. and {S{\'a}nchez-Janssen}, Rub{\'e}n and {Spengler}, Chelsea and {Toloba}, Elisa and {Zhang}, Hongxin and {Zhu}, Mingcheng},
        title = "{An evolutionary continuum from nucleated dwarf galaxies to star clusters}",
      journal = {\nat},
         year = 2023,
        month = nov,
       volume = {623},
       number = {7986},
        pages = {296-300},
          doi = {10.1038/s41586-023-06650-z},
archivePrefix = {arXiv},
       eprint = {2311.05448},
 primaryClass = {astro-ph.GA},
       adsurl = {https://ui.adsabs.harvard.edu/abs/2023Natur.623..296W}
}

@ARTICLE{Fahrion2024,
       author = {{Fahrion}, Katja and {B{\"o}ker}, Torsten and {Perna}, Michele and {Beck}, Tracy L. and {Maiolino}, Roberto and {Arribas}, Santiago and {Bunker}, Andrew J. and {Charlot}, Stephane and {Ceci}, Matteo and {Cresci}, Giovanni and {De Marchi}, Guido and {L{\"u}tzgendorf}, Nora and {Ulivi}, Lorenzo},
        title = "{Growing a nuclear star cluster from star formation and cluster mergers: The JWST NIRSpec view of NGC 4654}",
      journal = {\aap},
         year = 2024,
        month = jul,
       volume = {687},
          eid = {A83},
        pages = {A83},
          doi = {10.1051/0004-6361/202449629},
archivePrefix = {arXiv},
       eprint = {2404.08910},
 primaryClass = {astro-ph.GA},
       adsurl = {https://ui.adsabs.harvard.edu/abs/2024A&A...687A..83F}
}

@ARTICLE{Kacharov2018,
       author = {{Kacharov}, Nikolay and {Neumayer}, Nadine and {Seth}, Anil C. and {Cappellari}, Michele and {McDermid}, Richard and {Walcher}, C. Jakob and {B{\"o}ker}, Torsten},
        title = "{Stellar populations and star formation histories of the nuclear star clusters in six nearby galaxies}",
      journal = {\mnras},
         year = 2018,
        month = oct,
       volume = {480},
       number = {2},
        pages = {1973-1998},
          doi = {10.1093/mnras/sty1985},
archivePrefix = {arXiv},
       eprint = {1807.08765},
 primaryClass = {astro-ph.GA},
       adsurl = {https://ui.adsabs.harvard.edu/abs/2018MNRAS.480.1973K}
}

@ARTICLE{Pfeffer2016,
       author = {{Pfeffer}, J. and {Hilker}, M. and {Baumgardt}, H. and {Griffen}, B.~F.},
        title = "{Constraining ultracompact dwarf galaxy formation with galaxy clusters in the local universe}",
      journal = {\mnras},
         year = 2016,
        month = may,
       volume = {458},
       number = {3},
        pages = {2492-2508},
          doi = {10.1093/mnras/stw498},
archivePrefix = {arXiv},
       eprint = {1603.00032},
 primaryClass = {astro-ph.GA},
       adsurl = {https://ui.adsabs.harvard.edu/abs/2016MNRAS.458.2492P}
}

@ARTICLE{Paturel1992,
       author = {{Paturel}, G. and {Garnier}, R.},
        title = "{The use of velocity dispersion in globular clusters as a distance indicator.}",
      journal = {\aap},
         year = 1992,
        month = feb,
       volume = {254},
        pages = {93-95},
       adsurl = {https://ui.adsabs.harvard.edu/abs/1992A&A...254...93P}
}

@ARTICLE{Strader2009,
       author = {{Strader}, Jay and {Smith}, Graeme H. and {Larsen}, Soeren and {Brodie}, Jean P. and {Huchra}, John P.},
        title = "{Mass-to-Light Ratios for M31 Globular Clusters: Age Dating and a Surprising Metallicity Trend}",
      journal = {\aj},
         year = 2009,
        month = aug,
       volume = {138},
       number = {2},
        pages = {547-557},
          doi = {10.1088/0004-6256/138/2/547},
archivePrefix = {arXiv},
       eprint = {0906.0397},
 primaryClass = {astro-ph.GA},
       adsurl = {https://ui.adsabs.harvard.edu/abs/2009AJ....138..547S}
}

@ARTICLE{Baumgardt2020,
       author = {{Baumgardt}, H. and {Sollima}, A. and {Hilker}, M.},
        title = "{Absolute V-band magnitudes and mass-to-light ratios of Galactic globular clusters}",
      journal = {\pasa},
         year = 2020,
        month = nov,
       volume = {37},
          eid = {e046},
        pages = {e046},
          doi = {10.1017/pasa.2020.38},
archivePrefix = {arXiv},
       eprint = {2009.09611},
 primaryClass = {astro-ph.GA},
       adsurl = {https://ui.adsabs.harvard.edu/abs/2020PASA...37...46B}
}

@ARTICLE{Caldwell2011,
       author = {{Caldwell}, Nelson and {Schiavon}, Ricardo and {Morrison}, Heather and {Rose}, James A. and {Harding}, Paul},
        title = "{Star Clusters in M31. II. Old Cluster Metallicities and Ages from Hectospec Data}",
      journal = {\aj},
         year = 2011,
        month = feb,
       volume = {141},
       number = {2},
          eid = {61},
        pages = {61},
          doi = {10.1088/0004-6256/141/2/61},
archivePrefix = {arXiv},
       eprint = {1101.3779},
 primaryClass = {astro-ph.GA},
       adsurl = {https://ui.adsabs.harvard.edu/abs/2011AJ....141...61C}
}

@ARTICLE{Strader2011,
       author = {{Strader}, Jay and {Caldwell}, Nelson and {Seth}, Anil C.},
        title = "{Star Clusters in M31. V. Internal Dynamical Trends: Some Troublesome, Some Reassuring}",
      journal = {\aj},
         year = 2011,
        month = jul,
       volume = {142},
       number = {1},
          eid = {8},
        pages = {8},
          doi = {10.1088/0004-6256/142/1/8},
archivePrefix = {arXiv},
       eprint = {1104.4649},
 primaryClass = {astro-ph.CO},
       adsurl = {https://ui.adsabs.harvard.edu/abs/2011AJ....142....8S}
}

@ARTICLE{Hau2009,
       author = {{Hau}, George K.~T. and {Spitler}, Lee R. and {Forbes}, Duncan A. and {Proctor}, Robert N. and {Strader}, Jay and {Mendel}, J. Trevor and {Brodie}, Jean P. and {Harris}, William E.},
        title = "{An ultra-compact dwarf around the Sombrero galaxy (M104): the nearest massive UCD}",
      journal = {\mnras},
         year = 2009,
        month = mar,
       volume = {394},
       number = {1},
        pages = {L97-L101},
          doi = {10.1111/j.1745-3933.2009.00618.x},
archivePrefix = {arXiv},
       eprint = {0901.1693},
 primaryClass = {astro-ph.GA},
       adsurl = {https://ui.adsabs.harvard.edu/abs/2009MNRAS.394L..97H}
}

@ARTICLE{McLaughlin2000,
       author = {{McLaughlin}, Dean E.},
        title = "{Binding Energy and the Fundamental Plane of Globular Clusters}",
      journal = {\apj},
         year = 2000,
        month = aug,
       volume = {539},
       number = {2},
        pages = {618-640},
          doi = {10.1086/309247},
archivePrefix = {arXiv},
       eprint = {astro-ph/0002086},
 primaryClass = {astro-ph},
       adsurl = {https://ui.adsabs.harvard.edu/abs/2000ApJ...539..618M}
}

@ARTICLE{FaberJackson1976,
       author = {{Faber}, S.~M. and {Jackson}, R.~E.},
        title = "{Velocity dispersions and mass-to-light ratios for elliptical galaxies.}",
      journal = {\apj},
         year = 1976,
        month = mar,
       volume = {204},
        pages = {668-683},
          doi = {10.1086/154215},
       adsurl = {https://ui.adsabs.harvard.edu/abs/1976ApJ...204..668F}
}

@ARTICLE{Djorgovski1995,
       author = {{Djorgovski}, S.},
        title = "{The Fundamental Plane Correlations for Globular Clusters}",
      journal = {\apjl},
         year = 1995,
        month = jan,
       volume = {438},
        pages = {L29},
          doi = {10.1086/187707},
       adsurl = {https://ui.adsabs.harvard.edu/abs/1995ApJ...438L..29D}
}

@INPROCEEDINGS{Pryor1993,
       author = {{Pryor}, C. and {Meylan}, G.},
        title = "{Velocity Dispersions for Galactic Globular Clusters}",
    booktitle = {Structure and Dynamics of Globular Clusters},
         year = 1993,
       editor = {{Djorgovski}, S.~G. and {Meylan}, Georges},
       series = {Astronomical Society of the Pacific Conference Series},
       volume = {50},
        month = jan,
        pages = {357},
       adsurl = {https://ui.adsabs.harvard.edu/abs/1993ASPC...50..357P}
}

@ARTICLE{AlvesBrito2011,
       author = {{Alves-Brito}, Alan and {Hau}, George K.~T. and {Forbes}, Duncan A. and {Spitler}, Lee R. and {Strader}, Jay and {Brodie}, Jean P. and {Rhode}, Katherine L.},
        title = "{Spectra of globular clusters in the Sombrero galaxy: evidence for spectroscopic metallicity bimodality}",
      journal = {\mnras},
         year = 2011,
        month = nov,
       volume = {417},
       number = {3},
        pages = {1823-1838},
          doi = {10.1111/j.1365-2966.2011.19368.x},
archivePrefix = {arXiv},
       eprint = {1107.0757},
 primaryClass = {astro-ph.CO},
       adsurl = {https://ui.adsabs.harvard.edu/abs/2011MNRAS.417.1823A}
}

@ARTICLE{RhodeZepf2004,
       author = {{Rhode}, Katherine L. and {Zepf}, Stephen E.},
        title = "{The Globular Cluster Systems of the Early-Type Galaxies NGC 3379, NGC 4406, and NGC 4594 and Implications for Galaxy Formation}",
      journal = {\aj},
         year = 2004,
        month = jan,
       volume = {127},
       number = {1},
        pages = {302-317},
          doi = {10.1086/380616},
archivePrefix = {arXiv},
       eprint = {astro-ph/0310277},
 primaryClass = {astro-ph},
       adsurl = {https://ui.adsabs.harvard.edu/abs/2004AJ....127..302R}
}

@ARTICLE{HarrisHarrisHarris1984,
       author = {{Harris}, W.~E. and {Harris}, H.~C. and {Harris}, G.~L.~H.},
        title = "{Globular clusters in galaxies beyond the Local Group. III. NGC 4594 (The SOMBRERO).}",
      journal = {\aj},
         year = 1984,
        month = feb,
       volume = {89},
        pages = {216-223},
          doi = {10.1086/113504},
       adsurl = {https://ui.adsabs.harvard.edu/abs/1984AJ.....89..216H}
}

@ARTICLE{Wakamatsu1977,
       author = {{Wakamatsu}, K. -I.},
        title = "{Radial distribution and total number of globular clusters in M104.}",
      journal = {\pasp},
         year = 1977,
        month = jun,
       volume = {89},
        pages = {267-270},
          doi = {10.1086/130114},
       adsurl = {https://ui.adsabs.harvard.edu/abs/1977PASP...89..267W}
}

@INCOLLECTION{Beasley2020,
       author = {{Beasley}, Michael A.},
        title = "{Globular Cluster Systems and Galaxy Formation}",
    booktitle = {Reviews in Frontiers of Modern Astrophysics; From Space Debris to Cosmology},
         year = 2020,
        pages = {245-277},
          doi = {10.1007/978-3-030-38509-5\_9},
       adsurl = {https://ui.adsabs.harvard.edu/abs/2020rfma.book..245B}
}

@ARTICLE{Brodie2006,
       author = {{Brodie}, Jean P. and {Strader}, Jay},
        title = "{Extragalactic Globular Clusters and Galaxy Formation}",
      journal = {\araa},
         year = 2006,
        month = sep,
       volume = {44},
       number = {1},
        pages = {193-267},
          doi = {10.1146/annurev.astro.44.051905.092441},
archivePrefix = {arXiv},
       eprint = {astro-ph/0602601},
 primaryClass = {astro-ph},
       adsurl = {https://ui.adsabs.harvard.edu/abs/2006ARA&A..44..193B}
}

@ARTICLE{Vazdekis2016,
       author = {{Vazdekis}, A. and {Koleva}, M. and {Ricciardelli}, E. and {R{\"o}ck}, B. and {Falc{\'o}n-Barroso}, J.},
        title = "{UV-extended E-MILES stellar population models: young components in massive early-type galaxies}",
      journal = {\mnras},
         year = 2016,
        month = dec,
       volume = {463},
       number = {4},
        pages = {3409-3436},
          doi = {10.1093/mnras/stw2231},
archivePrefix = {arXiv},
       eprint = {1612.01187},
 primaryClass = {astro-ph.GA},
       adsurl = {https://ui.adsabs.harvard.edu/abs/2016MNRAS.463.3409V}
}

@ARTICLE{Vazdekis2010,
       author = {{Vazdekis}, A. and {S{\'a}nchez-Bl{\'a}zquez}, P. and {Falc{\'o}n-Barroso}, J. and {Cenarro}, A.~J. and {Beasley}, M.~A. and {Cardiel}, N. and {Gorgas}, J. and {Peletier}, R.~F.},
        title = "{Evolutionary stellar population synthesis with MILES - I. The base models and a new line index system}",
      journal = {\mnras},
         year = 2010,
        month = jun,
       volume = {404},
       number = {4},
        pages = {1639-1671},
          doi = {10.1111/j.1365-2966.2010.16407.x},
archivePrefix = {arXiv},
       eprint = {1004.4439},
 primaryClass = {astro-ph.CO},
       adsurl = {https://ui.adsabs.harvard.edu/abs/2010MNRAS.404.1639V}
}

@ARTICLE{Beasley2008,
       author = {{Beasley}, Michael A. and {Bridges}, Terry and {Peng}, Eric and {Harris}, William E. and {Harris}, Gretchen L.~H. and {Forbes}, Duncan A. and {Mackie}, Glen},
        title = "{A 2dF spectroscopic study of globular clusters in NGC 5128: probing the formation history of the nearest giant elliptical}",
      journal = {\mnras},
         year = 2008,
        month = may,
       volume = {386},
       number = {3},
        pages = {1443-1463},
          doi = {10.1111/j.1365-2966.2008.13123.x},
archivePrefix = {arXiv},
       eprint = {0803.1066},
 primaryClass = {astro-ph},
       adsurl = {https://ui.adsabs.harvard.edu/abs/2008MNRAS.386.1443B}
}

@ARTICLE{Beasley2025,
       author = {{Beasley}, M.~A. and {Fahrion}, K. and {Guerra Arencibia}, S. and {Gvozdenko}, A. and {Montes}, M.},
        title = "{A new way to measure the distance to NGC1052-DF2}",
      journal = {\aap},
         year = 2025,
        month = may,
       volume = {697},
          eid = {A144},
        pages = {A144},
          doi = {10.1051/0004-6361/202452446},
archivePrefix = {arXiv},
       eprint = {2503.03403},
 primaryClass = {astro-ph.GA},
       adsurl = {https://ui.adsabs.harvard.edu/abs/2025A&A...697A.144B}
}

@ARTICLE{Husser2013,
       author = {{Husser}, T. -O. and {Wende-von Berg}, S. and {Dreizler}, S. and {Homeier}, D. and {Reiners}, A. and {Barman}, T. and {Hauschildt}, P.~H.},
        title = "{A new extensive library of PHOENIX stellar atmospheres and synthetic spectra}",
      journal = {\aap},
         year = 2013,
        month = may,
       volume = {553},
          eid = {A6},
        pages = {A6},
          doi = {10.1051/0004-6361/201219058},
archivePrefix = {arXiv},
       eprint = {1303.5632},
 primaryClass = {astro-ph.SR},
       adsurl = {https://ui.adsabs.harvard.edu/abs/2013A&A...553A...6H}
}

@ARTICLE{Spitler2006,
       author = {{Spitler}, Lee R. and {Larsen}, S{\o}ren S. and {Strader}, Jay and {Brodie}, Jean P. and {Forbes}, Duncan A. and {Beasley}, Michael A.},
        title = "{Hubble Space Telescope ACS Wide-Field Photometry of the Sombrero Galaxy Globular Cluster System}",
      journal = {\aj},
         year = 2006,
        month = oct,
       volume = {132},
       number = {4},
        pages = {1593-1609},
          doi = {10.1086/507328},
archivePrefix = {arXiv},
       eprint = {astro-ph/0606337},
 primaryClass = {astro-ph},
       adsurl = {https://ui.adsabs.harvard.edu/abs/2006AJ....132.1593S}
}

@ARTICLE{Harris2010,
       author = {{Harris}, William E. and {Spitler}, Lee R. and {Forbes}, Duncan A. and {Bailin}, Jeremy},
        title = "{Diamonds on the Hat: globular clusters in the Sombrero galaxy (M104)}",
      journal = {\mnras},
         year = 2010,
        month = jan,
       volume = {401},
       number = {3},
        pages = {1965-1982},
          doi = {10.1111/j.1365-2966.2009.15783.x},
archivePrefix = {arXiv},
       eprint = {0909.4805},
 primaryClass = {astro-ph.GA},
       adsurl = {https://ui.adsabs.harvard.edu/abs/2010MNRAS.401.1965H}
}

@ARTICLE{Larsen2002,
       author = {{Larsen}, S{\o}ren S. and {Brodie}, Jean P. and {Beasley}, Michael A. and {Forbes}, Duncan A.},
        title = "{Keck Spectroscopy of Globular Clusters in the Sombrero Galaxy}",
      journal = {\aj},
         year = 2002,
        month = aug,
       volume = {124},
       number = {2},
        pages = {828-838},
          doi = {10.1086/341389},
archivePrefix = {arXiv},
       eprint = {astro-ph/0204420},
 primaryClass = {astro-ph},
       adsurl = {https://ui.adsabs.harvard.edu/abs/2002AJ....124..828L}
}

@ARTICLE{Bridges1997,
       author = {{Bridges}, T.~J. and {Ashman}, K.~M. and {Zepf}, S.~E. and {Carter}, D. and {Hanes}, D.~A. and {Sharples}, R.~M. and {Kavelaars}, J.~J.},
        title = "{Kinematics and metallicities of globular clusters in M104}",
      journal = {\mnras},
         year = 1997,
        month = jan,
       volume = {284},
       number = {2},
        pages = {376-384},
          doi = {10.1093/mnras/284.2.376},
       adsurl = {https://ui.adsabs.harvard.edu/abs/1997MNRAS.284..376B}
}

@ARTICLE{McQuinn2016,
       author = {{McQuinn}, Kristen. B.~W. and {Skillman}, Evan D. and {Dolphin}, Andrew E. and {Berg}, Danielle and {Kennicutt}, Robert},
        title = "{The Distance to M104}",
      journal = {\aj},
         year = 2016,
        month = nov,
       volume = {152},
       number = {5},
          eid = {144},
        pages = {144},
          doi = {10.3847/0004-6256/152/5/144},
       adsurl = {https://ui.adsabs.harvard.edu/abs/2016AJ....152..144M}
}

@ARTICLE{Cohen2020,
       author = {{Cohen}, Roger E. and {Goudfrooij}, Paul and {Correnti}, Matteo and {Gnedin}, Oleg Y. and {Harris}, William E. and {Chandar}, Rupali and {Puzia}, Thomas H. and {S{\'a}nchez-Janssen}, Rub{\'e}n},
        title = "{The Strikingly Metal-rich Halo of the Sombrero Galaxy}",
      journal = {\apj},
         year = 2020,
        month = feb,
       volume = {890},
       number = {1},
          eid = {52},
        pages = {52},
          doi = {10.3847/1538-4357/ab64e9},
archivePrefix = {arXiv},
       eprint = {2001.01670},
 primaryClass = {astro-ph.GA},
       adsurl = {https://ui.adsabs.harvard.edu/abs/2020ApJ...890...52C}
}

@ARTICLE{Dowell2014,
       author = {{Dowell}, Jessica L. and {Rhode}, Katherine L. and {Bridges}, Terry J. and {Zepf}, Stephen E. and {Gebhardt}, Karl and {Freeman}, Ken C. and {de Boer}, Elizabeth Wylie},
        title = "{Beyond the Brim of the Hat: Kinematics of Globular Clusters out to Large Radii in the Sombrero Galaxy}",
      journal = {\aj},
         year = 2014,
        month = jun,
       volume = {147},
       number = {6},
          eid = {150},
        pages = {150},
          doi = {10.1088/0004-6256/147/6/150},
archivePrefix = {arXiv},
       eprint = {1403.7227},
 primaryClass = {astro-ph.GA},
       adsurl = {https://ui.adsabs.harvard.edu/abs/2014AJ....147..150D}
}

@ARTICLE{Kang2022,
       author = {{Kang}, Jisu and {Lee}, Myung Gyoon and {Jang}, In Sung and {Ko}, Youkyung and {Sohn}, Jubee and {Hwang}, Narae and {Park}, Byeong-Gon},
        title = "{Tracing the Giant Outer Halo of the Mysterious Massive Disk Galaxy M104. I. Photometry of the Extended Globular Cluster Systems}",
      journal = {\apj},
         year = 2022,
        month = nov,
       volume = {939},
       number = {2},
          eid = {74},
        pages = {74},
          doi = {10.3847/1538-4357/ac9670},
archivePrefix = {arXiv},
       eprint = {2209.14677},
 primaryClass = {astro-ph.GA},
       adsurl = {https://ui.adsabs.harvard.edu/abs/2022ApJ...939...74K}
}

@ARTICLE{Emsellem1995,
       author = {{Emsellem}, E.},
        title = "{The Sombrero galaxy. I. Modelling the dust content.}",
      journal = {\aap},
         year = 1995,
        month = nov,
       volume = {303},
        pages = {673},
       adsurl = {https://ui.adsabs.harvard.edu/abs/1995A&A...303..673E}
}

@ARTICLE{Jardel2011,
       author = {{Jardel}, John R. and {Gebhardt}, Karl and {Shen}, Juntai and {Fisher}, David B. and {Kormendy}, John and {Kinzler}, Jeffry and {Lauer}, Tod R. and {Richstone}, Douglas and {G{\"u}ltekin}, K.},
        title = "{Orbit-based Dynamical Models of the Sombrero Galaxy (NGC 4594)}",
      journal = {\apj},
         year = 2011,
        month = sep,
       volume = {739},
       number = {1},
          eid = {21},
        pages = {21},
          doi = {10.1088/0004-637X/739/1/21},
archivePrefix = {arXiv},
       eprint = {1107.1238},
 primaryClass = {astro-ph.CO},
       adsurl = {https://ui.adsabs.harvard.edu/abs/2011ApJ...739...21J}
}

@ARTICLE{Tonry2001,
       author = {{Tonry}, John L. and {Dressler}, Alan and {Blakeslee}, John P. and {Ajhar}, Edward A. and {Fletcher}, Andr{\'e} B. and {Luppino}, Gerard A. and {Metzger}, Mark R. and {Moore}, Christopher B.},
        title = "{The SBF Survey of Galaxy Distances. IV. SBF Magnitudes, Colors, and Distances}",
      journal = {\apj},
         year = 2001,
        month = jan,
       volume = {546},
       number = {2},
        pages = {681-693},
          doi = {10.1086/318301},
archivePrefix = {arXiv},
       eprint = {astro-ph/0011223},
 primaryClass = {astro-ph},
       adsurl = {https://ui.adsabs.harvard.edu/abs/2001ApJ...546..681T}
}

@ARTICLE{Ferrarese2000,
       author = {{Ferrarese}, Laura and {Mould}, Jeremy R. and {Kennicutt}, Jr., Robert C. and {Huchra}, John and {Ford}, Holland C. and {Freedman}, Wendy L. and {Stetson}, Peter B. and {Madore}, Barry F. and {Sakai}, Shoko and {Gibson}, Brad K. and {Graham}, John A. and {Hughes}, Shaun M. and {Illingworth}, Garth D. and {Kelson}, Daniel D. and {Macri}, Lucas and {Sebo}, Kim and {Silbermann}, N.~A.},
        title = "{The Hubble Space Telescope Key Project on the Extragalactic Distance Scale. XXVI. The Calibration of Population II Secondary Distance Indicators and the Value of the Hubble Constant}",
      journal = {\apj},
         year = 2000,
        month = feb,
       volume = {529},
       number = {2},
        pages = {745-767},
          doi = {10.1086/308309},
archivePrefix = {arXiv},
       eprint = {astro-ph/9908192},
 primaryClass = {astro-ph},
       adsurl = {https://ui.adsabs.harvard.edu/abs/2000ApJ...529..745F}
}

@ARTICLE{Lotz2001,
       author = {{Lotz}, Jennifer M. and {Telford}, Rosemary and {Ferguson}, Henry C. and {Miller}, Bryan W. and {Stiavelli}, Massimo and {Mack}, Jennifer},
        title = "{Dynamical Friction in DE Globular Cluster Systems}",
      journal = {\apj},
         year = 2001,
        month = may,
       volume = {552},
       number = {2},
        pages = {572-581},
          doi = {10.1086/320545},
archivePrefix = {arXiv},
       eprint = {astro-ph/0102079},
 primaryClass = {astro-ph},
       adsurl = {https://ui.adsabs.harvard.edu/abs/2001ApJ...552..572L}
}

@ARTICLE{Tremaine1976,
       author = {{Tremaine}, S.~D.},
        title = "{The formation of the nuclei of galaxies. II. The local group.}",
      journal = {\apj},
         year = 1976,
        month = jan,
       volume = {203},
        pages = {345-351},
          doi = {10.1086/154085},
       adsurl = {https://ui.adsabs.harvard.edu/abs/1976ApJ...203..345T}
}

@ARTICLE{Turner2012,
       author = {{Turner}, Monica L. and {C{\^o}t{\'e}}, Patrick and {Ferrarese}, Laura and {Jord{\'a}n}, Andr{\'e}s and {Blakeslee}, John P. and {Mei}, Simona and {Peng}, Eric W. and {West}, Michael J.},
        title = "{The ACS Fornax Cluster Survey. VI. The Nuclei of Early-type Galaxies in the Fornax Cluster}",
      journal = {\apjs},
         year = 2012,
        month = nov,
       volume = {203},
       number = {1},
          eid = {5},
        pages = {5},
          doi = {10.1088/0067-0049/203/1/5},
archivePrefix = {arXiv},
       eprint = {1208.0338},
 primaryClass = {astro-ph.CO},
       adsurl = {https://ui.adsabs.harvard.edu/abs/2012ApJS..203....5T}
}

@ARTICLE{GierszHeggie1994,
       author = {{Giersz}, M. and {Heggie}, D.~C.},
        title = "{Statistics of N-Body Simulations - Part One - Equal Masses Before Core Collapse}",
      journal = {\mnras},
         year = 1994,
        month = may,
       volume = {268},
        pages = {257},
          doi = {10.1093/mnras/268.1.257},
archivePrefix = {arXiv},
       eprint = {astro-ph/9305008},
 primaryClass = {astro-ph},
       adsurl = {https://ui.adsabs.harvard.edu/abs/1994MNRAS.268..257G}
}

@ARTICLE{Barmby2007,
       author = {{Barmby}, Pauline and {McLaughlin}, Dean E. and {Harris}, William E. and {Harris}, Gretchen L.~H. and {Forbes}, Duncan A.},
        title = "{Structural Parameters for Globular Clusters in M31 and Generalizations for the Fundamental Plane}",
      journal = {\aj},
         year = 2007,
        month = jun,
       volume = {133},
       number = {6},
        pages = {2764-2786},
          doi = {10.1086/516777},
archivePrefix = {arXiv},
       eprint = {0704.2057},
 primaryClass = {astro-ph},
       adsurl = {https://ui.adsabs.harvard.edu/abs/2007AJ....133.2764B}
}

@ARTICLE{Peacock2010,
       author = {{Peacock}, Mark B. and {Maccarone}, Thomas J. and {Knigge}, Christian and {Kundu}, Arunav and {Waters}, Christopher Z. and {Zepf}, Stephen E. and {Zurek}, David R.},
        title = "{The M31 globular cluster system: ugriz and K-band photometry and structural parameters}",
      journal = {\mnras},
         year = 2010,
        month = feb,
       volume = {402},
       number = {2},
        pages = {803-818},
          doi = {10.1111/j.1365-2966.2009.15952.x},
archivePrefix = {arXiv},
       eprint = {0910.5475},
 primaryClass = {astro-ph.CO},
       adsurl = {https://ui.adsabs.harvard.edu/abs/2010MNRAS.402..803P}
}

@ARTICLE{Bridges1992,
       author = {{Bridges}, Terry J. and {Hanes}, David A.},
        title = "{The Globular Cluster System of NGC 4594 (The Sombrero)}",
      journal = {\aj},
         year = 1992,
        month = mar,
       volume = {103},
        pages = {800},
          doi = {10.1086/116102},
       adsurl = {https://ui.adsabs.harvard.edu/abs/1992AJ....103..800B}
}

@ARTICLE{Forbes1997,
       author = {{Forbes}, Duncan A. and {Grillmair}, Carl J. and {Smith}, R. Chris},
        title = "{Globular Clusters in the Sombrero Galaxy (NGC 4594)}",
      journal = {\aj},
         year = 1997,
        month = may,
       volume = {113},
        pages = {1648},
          doi = {10.1086/118381},
archivePrefix = {arXiv},
       eprint = {astro-ph/9701144},
 primaryClass = {astro-ph},
       adsurl = {https://ui.adsabs.harvard.edu/abs/1997AJ....113.1648F}
}

@ARTICLE{Bridges2007,
       author = {{Bridges}, Terry J. and {Rhode}, Katherine L. and {Zepf}, Stephen E. and {Freeman}, Ken C.},
        title = "{Spectroscopy of Globular Clusters out to Large Radius in the Sombrero Galaxy}",
      journal = {\apj},
         year = 2007,
        month = apr,
       volume = {658},
       number = {2},
        pages = {980-992},
          doi = {10.1086/511413},
archivePrefix = {arXiv},
       eprint = {astro-ph/0611896},
 primaryClass = {astro-ph},
       adsurl = {https://ui.adsabs.harvard.edu/abs/2007ApJ...658..980B}
}

@ARTICLE{Coelho2014,
       author = {{Coelho}, P.~R.~T.},
        title = "{A new library of theoretical stellar spectra with scaled-solar and {\ensuremath{\alpha}}-enhanced mixtures}",
      journal = {\mnras},
         year = 2014,
        month = may,
       volume = {440},
       number = {2},
        pages = {1027-1043},
          doi = {10.1093/mnras/stu365},
archivePrefix = {arXiv},
       eprint = {1404.3243},
 primaryClass = {astro-ph.SR},
       adsurl = {https://ui.adsabs.harvard.edu/abs/2014MNRAS.440.1027C}
}

@ARTICLE{Beasley2024,
       author = {{Beasley}, Michael A. and {Fahrion}, Katja and {Gvozdenko}, Anastasia},
        title = "{Measuring distances to galaxies with globular cluster velocity dispersions}",
      journal = {\mnras},
         year = 2024,
        month = jan,
       volume = {527},
       number = {3},
        pages = {5767-5775},
          doi = {10.1093/mnras/stad3541},
archivePrefix = {arXiv},
       eprint = {2312.01420},
 primaryClass = {astro-ph.GA},
       adsurl = {https://ui.adsabs.harvard.edu/abs/2024MNRAS.527.5767B}
}

@ARTICLE{BaumgardtHilker2018,
       author = {{Baumgardt}, H. and {Hilker}, M.},
        title = "{A catalogue of masses, structural parameters, and velocity dispersion profiles of 112 Milky Way globular clusters}",
      journal = {\mnras},
         year = 2018,
        month = aug,
       volume = {478},
       number = {2},
        pages = {1520-1557},
          doi = {10.1093/mnras/sty1057},
archivePrefix = {arXiv},
       eprint = {1804.08359},
 primaryClass = {astro-ph.GA},
       adsurl = {https://ui.adsabs.harvard.edu/abs/2018MNRAS.478.1520B}
}

@software{synphot,
       author = {{STScI Development Team}},
        title = "{synphot: Synthetic photometry using Astropy}",
 howpublished = {Astrophysics Source Code Library, record ascl:1811.001},
         year = 2018,
        month = nov,
          eid = {ascl:1811.001},
       adsurl = {https://ui.adsabs.harvard.edu/abs/2018ascl.soft11001S}
}

@ARTICLE{emcee,
       author = {{Foreman-Mackey}, Daniel and {Hogg}, David W. and {Lang}, Dustin and {Goodman}, Jonathan},
        title = "{emcee: The MCMC Hammer}",
      journal = {\pasp},
         year = 2013,
        month = mar,
       volume = {125},
       number = {925},
        pages = {306},
          doi = {10.1086/670067},
archivePrefix = {arXiv},
       eprint = {1202.3665},
 primaryClass = {astro-ph.IM},
       adsurl = {https://ui.adsabs.harvard.edu/abs/2013PASP..125..306F}
}

@ARTICLE{Cappellari2023,
       author = {{Cappellari}, Michele},
        title = "{Full spectrum fitting with photometry in PPXF: stellar population versus dynamical masses, non-parametric star formation history and metallicity for 3200 LEGA-C galaxies at redshift z {\ensuremath{\approx}} 0.8}",
      journal = {\mnras},
         year = 2023,
        month = dec,
       volume = {526},
       number = {3},
        pages = {3273-3300},
          doi = {10.1093/mnras/stad2597},
archivePrefix = {arXiv},
       eprint = {2208.14974},
 primaryClass = {astro-ph.GA},
       adsurl = {https://ui.adsabs.harvard.edu/abs/2023MNRAS.526.3273C}
}

@BOOK{BinneyTremaine1987,
       author = {{Binney}, James and {Tremaine}, Scott},
        title = "{Galactic dynamics}",
         year = 1987,
       adsurl = {https://ui.adsabs.harvard.edu/abs/1987gady.book.....B}
}

@ARTICLE{Cappellari2004,
       author = {{Cappellari}, Michele and {Emsellem}, Eric},
        title = "{Parametric Recovery of Line-of-Sight Velocity Distributions from Absorption-Line Spectra of Galaxies via Penalized Likelihood}",
      journal = {\pasp},
         year = 2004,
        month = feb,
       volume = {116},
       number = {816},
        pages = {138-147},
          doi = {10.1086/381875},
archivePrefix = {arXiv},
       eprint = {astro-ph/0312201},
 primaryClass = {astro-ph},
       adsurl = {https://ui.adsabs.harvard.edu/abs/2004PASP..116..138C}
}

@ARTICLE{Cappellari2017,
       author = {{Cappellari}, Michele},
        title = "{Improving the full spectrum fitting method: accurate convolution with Gauss-Hermite functions}",
      journal = {\mnras},
         year = 2017,
        month = apr,
       volume = {466},
       number = {1},
        pages = {798-811},
          doi = {10.1093/mnras/stw3020},
archivePrefix = {arXiv},
       eprint = {1607.08538},
 primaryClass = {astro-ph.GA},
       adsurl = {https://ui.adsabs.harvard.edu/abs/2017MNRAS.466..798C}
}

@ARTICLE{Gadotti2012,
       author = {{Gadotti}, Dimitri A. and {S{\'a}nchez-Janssen}, Rub{\'e}n.},
        title = "{Surprises in image decomposition of edge-on galaxies: does Sombrero have a (classical) bulge?}",
      journal = {\mnras},
         year = 2012,
        month = jun,
       volume = {423},
       number = {1},
        pages = {877-888},
          doi = {10.1111/j.1365-2966.2012.20925.x},
archivePrefix = {arXiv},
       eprint = {1101.2900},
 primaryClass = {astro-ph.CO},
       adsurl = {https://ui.adsabs.harvard.edu/abs/2012MNRAS.423..877G}
}

@ARTICLE{Fahrion2021,
       author = {{Fahrion}, K. and {Lyubenova}, M. and {van de Ven}, G. and {Hilker}, M. and {Leaman}, R. and {Falc{\'o}n-Barroso}, J. and {Bittner}, A. and {Coccato}, L. and {Corsini}, E.~M. and {Gadotti}, D.~A. and {Iodice}, E. and {McDermid}, R.~M. and {Mart{\'\i}n-Navarro}, I. and {Pinna}, F. and {Poci}, A. and {Sarzi}, M. and {de Zeeuw}, P.~T. and {Zhu}, L.},
        title = "{Diversity of nuclear star cluster formation mechanisms revealed by their star formation histories}",
      journal = {\aap},
         year = 2021,
        month = jun,
       volume = {650},
          eid = {A137},
        pages = {A137},
          doi = {10.1051/0004-6361/202140644},
archivePrefix = {arXiv},
       eprint = {2104.06412},
 primaryClass = {astro-ph.GA},
       adsurl = {https://ui.adsabs.harvard.edu/abs/2021A&A...650A.137F}
}

@ARTICLE{astropy2013,
       author = {{Astropy Collaboration} and {Robitaille}, Thomas P. and {Tollerud}, Erik J. and {Greenfield}, Perry and {Droettboom}, Michael and {Bray}, Erik and {Aldcroft}, Tom and {Davis}, Matt and {Ginsburg}, Adam and {Price-Whelan}, Adrian M. and {Kerzendorf}, Wolfgang E. and {Conley}, Alexander and {Crighton}, Neil and {Barbary}, Kyle and {Muna}, Demitri and {Ferguson}, Henry and {Grollier}, Fr{\'e}d{\'e}ric and {Parikh}, Madhura M. and {Nair}, Prasanth H. and {Unther}, Hans M. and {Deil}, Christoph and {Woillez}, Julien and {Conseil}, Simon and {Kramer}, Roban and {Turner}, James E.~H. and {Singer}, Leo and {Fox}, Ryan and {Weaver}, Benjamin A. and {Zabalza}, Victor and {Edwards}, Zachary I. and {Azalee Bostroem}, K. and {Burke}, D.~J. and {Casey}, Andrew R. and {Crawford}, Steven M. and {Dencheva}, Nadia and {Ely}, Justin and {Jenness}, Tim and {Labrie}, Kathleen and {Lim}, Pey Lian and {Pierfederici}, Francesco and {Pontzen}, Andrew and {Ptak}, Andy and {Refsdal}, Brian and {Servillat}, Mathieu and {Streicher}, Ole},
        title = "{Astropy: A community Python package for astronomy}",
      journal = {\aap},
         year = 2013,
        month = oct,
       volume = {558},
          eid = {A33},
        pages = {A33},
          doi = {10.1051/0004-6361/201322068},
archivePrefix = {arXiv},
       eprint = {1307.6212},
 primaryClass = {astro-ph.IM},
       adsurl = {https://ui.adsabs.harvard.edu/abs/2013A&A...558A..33A}
}

@ARTICLE{astropy2018,
       author = {{Astropy Collaboration} and {Price-Whelan}, A.~M. and {Sip{\H{o}}cz}, B.~M. and {G{\"u}nther}, H.~M. and {Lim}, P.~L. and {Crawford}, S.~M. and {Conseil}, S. and {Shupe}, D.~L. and {Craig}, M.~W. and {Dencheva}, N. and {Ginsburg}, A. and {VanderPlas}, J.~T. and {Bradley}, L.~D. and {P{\'e}rez-Su{\'a}rez}, D. and {de Val-Borro}, M. and {Aldcroft}, T.~L. and {Cruz}, K.~L. and {Robitaille}, T.~P. and {Tollerud}, E.~J. and {Ardelean}, C. and {Babej}, T. and {Bach}, Y.~P. and {Bachetti}, M. and {Bakanov}, A.~V. and {Bamford}, S.~P. and {Barentsen}, G. and {Barmby}, P. and {Baumbach}, A. and {Berry}, K.~L. and {Biscani}, F. and {Boquien}, M. and {Bostroem}, K.~A. and {Bouma}, L.~G. and {Brammer}, G.~B. and {Bray}, E.~M. and {Breytenbach}, H. and {Buddelmeijer}, H. and {Burke}, D.~J. and {Calderone}, G. and {Cano Rodr{\'\i}guez}, J.~L. and {Cara}, M. and {Cardoso}, J.~V.~M. and {Cheedella}, S. and {Copin}, Y. and {Corrales}, L. and {Crichton}, D. and {D'Avella}, D. and {Deil}, C. and {Depagne}, {\'E}. and {Dietrich}, J.~P. and {Donath}, A. and {Droettboom}, M. and {Earl}, N. and {Erben}, T. and {Fabbro}, S. and {Ferreira}, L.~A. and {Finethy}, T. and {Fox}, R.~T. and {Garrison}, L.~H. and {Gibbons}, S.~L.~J. and {Goldstein}, D.~A. and {Gommers}, R. and {Greco}, J.~P. and {Greenfield}, P. and {Groener}, A.~M. and {Grollier}, F. and {Hagen}, A. and {Hirst}, P. and {Homeier}, D. and {Horton}, A.~J. and {Hosseinzadeh}, G. and {Hu}, L. and {Hunkeler}, J.~S. and {Ivezi{\'c}}, {\v{Z}}. and {Jain}, A. and {Jenness}, T. and {Kanarek}, G. and {Kendrew}, S. and {Kern}, N.~S. and {Kerzendorf}, W.~E. and {Khvalko}, A. and {King}, J. and {Kirkby}, D. and {Kulkarni}, A.~M. and {Kumar}, A. and {Lee}, A. and {Lenz}, D. and {Littlefair}, S.~P. and {Ma}, Z. and {Macleod}, D.~M. and {Mastropietro}, M. and {McCully}, C. and {Montagnac}, S. and {Morris}, B.~M. and {Mueller}, M. and {Mumford}, S.~J. and {Muna}, D. and {Murphy}, N.~A. and {Nelson}, S. and {Nguyen}, G.~H. and {Ninan}, J.~P. and {N{\"o}the}, M. and {Ogaz}, S. and {Oh}, S. and {Parejko}, J.~K. and {Parley}, N. and {Pascual}, S. and {Patil}, R. and {Patil}, A.~A. and {Plunkett}, A.~L. and {Prochaska}, J.~X. and {Rastogi}, T. and {Reddy Janga}, V. and {Sabater}, J. and {Sakurikar}, P. and {Seifert}, M. and {Sherbert}, L.~E. and {Sherwood-Taylor}, H. and {Shih}, A.~Y. and {Sick}, J. and {Silbiger}, M.~T. and {Singanamalla}, S. and {Singer}, L.~P. and {Sladen}, P.~H. and {Sooley}, K.~A. and {Sornarajah}, S. and {Streicher}, O. and {Teuben}, P. and {Thomas}, S.~W. and {Tremblay}, G.~R. and {Turner}, J.~E.~H. and {Terr{\'o}n}, V. and {van Kerkwijk}, M.~H. and {de la Vega}, A. and {Watkins}, L.~L. and {Weaver}, B.~A. and {Whitmore}, J.~B. and {Woillez}, J. and {Zabalza}, V. and {Astropy Contributors}},
        title = "{The Astropy Project: Building an Open-science Project and Status of the v2.0 Core Package}",
      journal = {\aj},
         year = 2018,
        month = sep,
       volume = {156},
       number = {3},
          eid = {123},
        pages = {123},
          doi = {10.3847/1538-3881/aabc4f},
archivePrefix = {arXiv},
       eprint = {1801.02634},
 primaryClass = {astro-ph.IM},
       adsurl = {https://ui.adsabs.harvard.edu/abs/2018AJ....156..123A}
}

@ARTICLE{astropy2022,
       author = {{Astropy Collaboration} and {Price-Whelan}, Adrian M. and {Lim}, Pey Lian and {Earl}, Nicholas and {Starkman}, Nathaniel and {Bradley}, Larry and {Shupe}, David L. and {Patil}, Aarya A. and {Corrales}, Lia and {Brasseur}, C.~E. and {N{\"o}the}, Maximilian and {Donath}, Axel and {Tollerud}, Erik and {Morris}, Brett M. and {Ginsburg}, Adam and {Vaher}, Eero and {Weaver}, Benjamin A. and {Tocknell}, James and {Jamieson}, William and {van Kerkwijk}, Marten H. and {Robitaille}, Thomas P. and {Merry}, Bruce and {Bachetti}, Matteo and {G{\"u}nther}, H. Moritz and {Aldcroft}, Thomas L. and {Alvarado-Montes}, Jaime A. and {Archibald}, Anne M. and {B{\'o}di}, Attila and {Bapat}, Shreyas and {Barentsen}, Geert and {Baz{\'a}n}, Juanjo and {Biswas}, Manish and {Boquien}, M{\'e}d{\'e}ric and {Burke}, D.~J. and {Cara}, Daria and {Cara}, Mihai and {Conroy}, Kyle E. and {Conseil}, Simon and {Craig}, Matthew W. and {Cross}, Robert M. and {Cruz}, Kelle L. and {D'Eugenio}, Francesco and {Dencheva}, Nadia and {Devillepoix}, Hadrien A.~R. and {Dietrich}, J{\"o}rg P. and {Eigenbrot}, Arthur Davis and {Erben}, Thomas and {Ferreira}, Leonardo and {Foreman-Mackey}, Daniel and {Fox}, Ryan and {Freij}, Nabil and {Garg}, Suyog and {Geda}, Robel and {Glattly}, Lauren and {Gondhalekar}, Yash and {Gordon}, Karl D. and {Grant}, David and {Greenfield}, Perry and {Groener}, Austen M. and {Guest}, Steve and {Gurovich}, Sebastian and {Handberg}, Rasmus and {Hart}, Akeem and {Hatfield-Dodds}, Zac and {Homeier}, Derek and {Hosseinzadeh}, Griffin and {Jenness}, Tim and {Jones}, Craig K. and {Joseph}, Prajwel and {Kalmbach}, J. Bryce and {Karamehmetoglu}, Emir and {Ka{\l}uszy{\'n}ski}, Miko{\l}aj and {Kelley}, Michael S.~P. and {Kern}, Nicholas and {Kerzendorf}, Wolfgang E. and {Koch}, Eric W. and {Kulumani}, Shankar and {Lee}, Antony and {Ly}, Chun and {Ma}, Zhiyuan and {MacBride}, Conor and {Maljaars}, Jakob M. and {Muna}, Demitri and {Murphy}, N.~A. and {Norman}, Henrik and {O'Steen}, Richard and {Oman}, Kyle A. and {Pacifici}, Camilla and {Pascual}, Sergio and {Pascual-Granado}, J. and {Patil}, Rohit R. and {Perren}, Gabriel I. and {Pickering}, Timothy E. and {Rastogi}, Tanuj and {Roulston}, Benjamin R. and {Ryan}, Daniel F. and {Rykoff}, Eli S. and {Sabater}, Jose and {Sakurikar}, Parikshit and {Salgado}, Jes{\'u}s and {Sanghi}, Aniket and {Saunders}, Nicholas and {Savchenko}, Volodymyr and {Schwardt}, Ludwig and {Seifert-Eckert}, Michael and {Shih}, Albert Y. and {Jain}, Anany Shrey and {Shukla}, Gyanendra and {Sick}, Jonathan and {Simpson}, Chris and {Singanamalla}, Sudheesh and {Singer}, Leo P. and {Singhal}, Jaladh and {Sinha}, Manodeep and {Sip{\H{o}}cz}, Brigitta M. and {Spitler}, Lee R. and {Stansby}, David and {Streicher}, Ole and {{\v{S}}umak}, Jani and {Swinbank}, John D. and {Taranu}, Dan S. and {Tewary}, Nikita and {Tremblay}, Grant R. and {de Val-Borro}, Miguel and {Van Kooten}, Samuel J. and {Vasovi{\'c}}, Zlatan and {Verma}, Shresth and {de Miranda Cardoso}, Jos{\'e} Vin{\'\i}cius and {Williams}, Peter K.~G. and {Wilson}, Tom J. and {Winkel}, Benjamin and {Wood-Vasey}, W.~M. and {Xue}, Rui and {Yoachim}, Peter and {Zhang}, Chen and {Zonca}, Andrea and {Astropy Project Contributors}},
        title = "{The Astropy Project: Sustaining and Growing a Community-oriented Open-source Project and the Latest Major Release (v5.0) of the Core Package}",
      journal = {\apj},
         year = 2022,
        month = aug,
       volume = {935},
       number = {2},
          eid = {167},
        pages = {167},
          doi = {10.3847/1538-4357/ac7c74},
archivePrefix = {arXiv},
       eprint = {2206.14220},
 primaryClass = {astro-ph.IM},
       adsurl = {https://ui.adsabs.harvard.edu/abs/2022ApJ...935..167A}
}

@ARTICLE{Karunakaran2020,
       author = {{Karunakaran}, Ananthan and {Spekkens}, Kristine and {Zaritsky}, Dennis and {Donnerstein}, Richard L. and {Kadowaki}, Jennifer and {Dey}, Arjun},
        title = "{Systematically Measuring Ultradiffuse Galaxies in H I: Results from the Pilot Survey}",
      journal = {\apj},
         year = 2020,
        month = oct,
       volume = {902},
       number = {1},
          eid = {39},
        pages = {39},
          doi = {10.3847/1538-4357/abb464},
archivePrefix = {arXiv},
       eprint = {2005.14202},
 primaryClass = {astro-ph.GA},
       adsurl = {https://ui.adsabs.harvard.edu/abs/2020ApJ...902...39K}
}

@ARTICLE{Ciardullo1993,
       author = {{Ciardullo}, Robin and {Jacoby}, George H. and {Tonry}, John L.},
        title = "{A Comparison of the Planetary Nebula Luminosity Function and Surface Brightness Fluctuation Distance Scales}",
      journal = {\apj},
         year = 1993,
        month = dec,
       volume = {419},
        pages = {479},
          doi = {10.1086/173501},
       adsurl = {https://ui.adsabs.harvard.edu/abs/1993ApJ...419..479C}
}

@ARTICLE{Masters2010,
       author = {{Masters}, Karen L. and {Jord{\'a}n}, Andr{\'e}s and {C{\^o}t{\'e}}, Patrick and {Ferrarese}, Laura and {Blakeslee}, John P. and {Infante}, Leopoldo and {Peng}, Eric W. and {Mei}, Simona and {West}, Michael J.},
        title = "{The Advanced Camera for Surveys Fornax Cluster Survey. VII. Half-light Radii of Globular Clusters in Early-type Galaxies}",
      journal = {\apj},
         year = 2010,
        month = jun,
       volume = {715},
       number = {2},
        pages = {1419-1437},
          doi = {10.1088/0004-637X/715/2/1419},
archivePrefix = {arXiv},
       eprint = {1003.3450},
 primaryClass = {astro-ph.CO},
       adsurl = {https://ui.adsabs.harvard.edu/abs/2010ApJ...715.1419M}
}

@ARTICLE{Mieske2008,
       author = {{Mieske}, S. and {Hilker}, M. and {Jord{\'a}n}, A. and {Infante}, L. and {Kissler-Patig}, M. and {Rejkuba}, M. and {Richtler}, T. and {C{\^o}t{\'e}}, P. and {Baumgardt}, H. and {West}, M.~J. and {Ferrarese}, L. and {Peng}, E.~W.},
        title = "{The nature of UCDs: Internal dynamics from an expanded sample and homogeneous database}",
      journal = {\aap},
         year = 2008,
        month = sep,
       volume = {487},
       number = {3},
        pages = {921-935},
          doi = {10.1051/0004-6361:200810077},
archivePrefix = {arXiv},
       eprint = {0806.0374},
 primaryClass = {astro-ph},
       adsurl = {https://ui.adsabs.harvard.edu/abs/2008A&A...487..921M}
}

@ARTICLE{Rejkuba2007,
       author = {{Rejkuba}, M. and {Dubath}, P. and {Minniti}, D. and {Meylan}, G.},
        title = "{Bright globular clusters in NGC 5128: the missing link between young massive clusters and evolved massive objects}",
      journal = {\aap},
         year = 2007,
        month = jul,
       volume = {469},
       number = {1},
        pages = {147-162},
          doi = {10.1051/0004-6361:20066493},
archivePrefix = {arXiv},
       eprint = {astro-ph/0703385},
 primaryClass = {astro-ph},
       adsurl = {https://ui.adsabs.harvard.edu/abs/2007A&A...469..147R}
}

@ARTICLE{Anand2024,
       author = {{Anand}, Gagandeep S. and {Tully}, R. Brent and {Cohen}, Yotam and {Makarov}, Dmitry I. and {Makarova}, Lidia N. and {Jensen}, Joseph B. and {Blakeslee}, John P. and {Cantiello}, Michele and {Kourkchi}, Ehsan and {Raimondo}, Gabriella},
        title = "{The TRGB-SBF Project. I. A Tip of the Red Giant Branch Distance to the Fornax Cluster with JWST}",
      journal = {\apj},
         year = 2024,
        month = oct,
       volume = {973},
       number = {2},
          eid = {83},
        pages = {83},
          doi = {10.3847/1538-4357/ad64c7},
archivePrefix = {arXiv},
       eprint = {2405.03743},
 primaryClass = {astro-ph.CO},
       adsurl = {https://ui.adsabs.harvard.edu/abs/2024ApJ...973...83A}
}

@ARTICLE{Jensen2025,
       author = {{Jensen}, Joseph B. and {Blakeslee}, John P. and {Cantiello}, Michele and {Cowles}, Mikaela and {Anand}, Gagandeep S. and {Tully}, R. Brent and {Kourkchi}, Ehsan and {Raimondo}, Gabriella},
        title = "{The TRGB‑SBF Project. III. Refining the HST Surface Brightness Fluctuation Distance Scale Calibration with JWST}",
      journal = {\apj},
         year = 2025,
        month = jul,
       volume = {987},
       number = {1},
          eid = {87},
        pages = {87},
          doi = {10.3847/1538-4357/addfd6},
archivePrefix = {arXiv},
       eprint = {2502.15935},
 primaryClass = {astro-ph.CO},
       adsurl = {https://ui.adsabs.harvard.edu/abs/2025ApJ...987...87J}
}

@ARTICLE{Carretta2000,
       author = {{Carretta}, Eugenio and {Gratton}, Raffaele G. and {Clementini}, Gisella and {Fusi Pecci}, Flavio},
        title = "{Distances, Ages, and Epoch of Formation of Globular Clusters}",
      journal = {\apj},
         year = 2000,
        month = apr,
       volume = {533},
       number = {1},
        pages = {215-235},
          doi = {10.1086/308629},
archivePrefix = {arXiv},
       eprint = {astro-ph/9902086},
 primaryClass = {astro-ph},
       adsurl = {https://ui.adsabs.harvard.edu/abs/2000ApJ...533..215C}
}

@ARTICLE{Rizzi2007,
       author = {{Rizzi}, Luca and {Tully}, R. Brent and {Makarov}, Dmitry and {Makarova}, Lidia and {Dolphin}, Andrew E. and {Sakai}, Shoko and {Shaya}, Edward J.},
        title = "{Tip of the Red Giant Branch Distances. II. Zero-Point Calibration}",
      journal = {\apj},
         year = 2007,
        month = jun,
       volume = {661},
       number = {2},
        pages = {815-829},
          doi = {10.1086/516566},
archivePrefix = {arXiv},
       eprint = {astro-ph/0701518},
 primaryClass = {astro-ph},
       adsurl = {https://ui.adsabs.harvard.edu/abs/2007ApJ...661..815R}
}

@ARTICLE{Villegas2010,
       author = {{Villegas}, Daniela and {Jord{\'a}n}, Andr{\'e}s and {Peng}, Eric W. and {Blakeslee}, John P. and {C{\^o}t{\'e}}, Patrick and {Ferrarese}, Laura and {Kissler-Patig}, Markus and {Mei}, Simona and {Infante}, Leopoldo and {Tonry}, John L. and {West}, Michael J.},
        title = "{The ACS Fornax Cluster Survey. VIII. The Luminosity Function of Globular Clusters in Virgo and Fornax Early-type Galaxies and Its Use as a Distance Indicator}",
      journal = {\apj},
         year = 2010,
        month = jul,
       volume = {717},
       number = {2},
        pages = {603-616},
          doi = {10.1088/0004-637X/717/2/603},
archivePrefix = {arXiv},
       eprint = {1004.2883},
 primaryClass = {astro-ph.CO},
       adsurl = {https://ui.adsabs.harvard.edu/abs/2010ApJ...717..603V}
}

@ARTICLE{Jensen2003,
       author = {{Jensen}, Joseph B. and {Tonry}, John L. and {Barris}, Brian J. and {Thompson}, Rodger I. and {Liu}, Michael C. and {Rieke}, Marcia J. and {Ajhar}, Edward A. and {Blakeslee}, John P.},
        title = "{Measuring Distances and Probing the Unresolved Stellar Populations of Galaxies Using Infrared Surface Brightness Fluctuations}",
      journal = {\apj},
         year = 2003,
        month = feb,
       volume = {583},
       number = {2},
        pages = {712-726},
          doi = {10.1086/345430},
archivePrefix = {arXiv},
       eprint = {astro-ph/0210129},
 primaryClass = {astro-ph},
       adsurl = {https://ui.adsabs.harvard.edu/abs/2003ApJ...583..712J}
}

@ARTICLE{Bottinelli1984,
       author = {{Bottinelli}, L. and {Gouguenheim}, L. and {Paturel}, G. and {de Vaucouleurs}, G.},
        title = "{HI line studies of galaxies. III. Distance moduli of 822 disk galaxies.}",
      journal = {\aaps},
         year = 1984,
        month = jun,
       volume = {56},
        pages = {381-413},
       adsurl = {https://ui.adsabs.harvard.edu/abs/1984A&AS...56..381B}
}

@BOOK{Tully1988,
       author = {{Tully}, R. Brent and {Fisher}, J. Richard},
        title = "{Catalog of Nearby Galaxies}",
         year = 1988,
       adsurl = {https://ui.adsabs.harvard.edu/abs/1988cng..book.....T}
}

@ARTICLE{Ajhar1997,
       author = {{Ajhar}, Edward A. and {Lauer}, Tod R. and {Tonry}, John L. and {Blakeslee}, John P. and {Dressler}, Alan and {Holtzman}, Jon A. and {Postman}, Marc},
        title = "{Calibration of the Surface Brightness Fluctuation Method for use with the Hubble Space Telescope.}",
      journal = {\aj},
         year = 1997,
        month = aug,
       volume = {114},
        pages = {626-634},
          doi = {10.1086/118498},
       adsurl = {https://ui.adsabs.harvard.edu/abs/1997AJ....114..626A}
}

@ARTICLE{Fahrion2025b,
       author = {{Fahrion}, Katja and {Beasley}, Michael A. and {Emsellem}, Eric and {Gvozdenko}, Anastasia and {M{\"u}ller}, Oliver and {Rejkuba}, Marina},
        title = "{Globular clusters in M104: Tracing kinematics and metallicities from the centre to the halo}",
      journal = {arXiv e-prints},
         year = 2025,
        month = aug,
          eid = {arXiv:2508.10100},
        pages = {arXiv:2508.10100},
          doi = {10.48550/arXiv.2508.10100},
archivePrefix = {arXiv},
       eprint = {2508.10100},
 primaryClass = {astro-ph.GA},
       adsurl = {https://ui.adsabs.harvard.edu/abs/2025arXiv250810100F}
}

@ARTICLE{Ford1996,
       author = {{Ford}, H.~C. and {Hui}, X. and {Ciardullo}, R. and {Jacoby}, G.~H. and {Freeman}, K.~C.},
        title = "{The Stellar Halo of M104. I. A Survey for Planetary Nebulae and the Planetary Nebula Luminosity Function Distance}",
      journal = {\apj},
         year = 1996,
        month = feb,
       volume = {458},
        pages = {455},
          doi = {10.1086/176828},
       adsurl = {https://ui.adsabs.harvard.edu/abs/1996ApJ...458..455F}
}

@ARTICLE{Sabbi2018,
       author = {{Sabbi}, E. and {Calzetti}, D. and {Ubeda}, L. and {Adamo}, A. and {Cignoni}, M. and {Thilker}, D. and {Aloisi}, A. and {Elmegreen}, B.~G. and {Elmegreen}, D.~M. and {Gouliermis}, D.~A. and {Grebel}, E.~K. and {Messa}, M. and {Smith}, L.~J. and {Tosi}, M. and {Dolphin}, A. and {Andrews}, J.~E. and {Ashworth}, G. and {Bright}, S.~N. and {Brown}, T.~M. and {Chandar}, R. and {Christian}, C. and {Clayton}, G.~C. and {Cook}, D.~O. and {Dale}, D.~A. and {de Mink}, S.~E. and {Dobbs}, C. and {Evans}, A.~S. and {Fumagalli}, M. and {Gallagher}, III, J.~S. and {Grasha}, K. and {Herrero}, A. and {Hunter}, D.~A. and {Johnson}, K.~E. and {Kahre}, L. and {Kennicutt}, R.~C. and {Kim}, H. and {Krumholz}, M.~R. and {Lee}, J.~C. and {Lennon}, D. and {Martin}, C. and {Nair}, P. and {Nota}, A. and {{\"O}stlin}, G. and {Pellerin}, A. and {Prieto}, J. and {Regan}, M.~W. and {Ryon}, J.~E. and {Sacchi}, E. and {Schaerer}, D. and {Schiminovich}, D. and {Shabani}, F. and {Van Dyk}, S.~D. and {Walterbos}, R. and {Whitmore}, B.~C. and {Wofford}, A.},
        title = "{The Resolved Stellar Populations in the LEGUS Galaxies1}",
      journal = {\apjs},
         year = 2018,
        month = mar,
       volume = {235},
       number = {1},
          eid = {23},
        pages = {23},
          doi = {10.3847/1538-4365/aaa8e5},
archivePrefix = {arXiv},
       eprint = {1801.05467},
 primaryClass = {astro-ph.GA},
       adsurl = {https://ui.adsabs.harvard.edu/abs/2018ApJS..235...23S}
}

@ARTICLE{Sorce2014,
       author = {{Sorce}, J.~G. and {Tully}, R.~B. and {Courtois}, H.~M. and {Jarrett}, T.~H. and {Neill}, J.~D. and {Shaya}, E.~J.},
        title = "{From Spitzer Galaxy photometry to Tully-Fisher distances}",
      journal = {\mnras},
         year = 2014,
        month = oct,
       volume = {444},
       number = {1},
        pages = {527-541},
          doi = {10.1093/mnras/stu1450},
archivePrefix = {arXiv},
       eprint = {1408.0729},
 primaryClass = {astro-ph.GA},
       adsurl = {https://ui.adsabs.harvard.edu/abs/2014MNRAS.444..527S}
}

 \begin{appendix}
\section{Radial trends of magnitudes and dispersions}
\label{ap:radial_trends}
\begin{figure}
    \centering
    \includegraphics[width=0.48\textwidth]{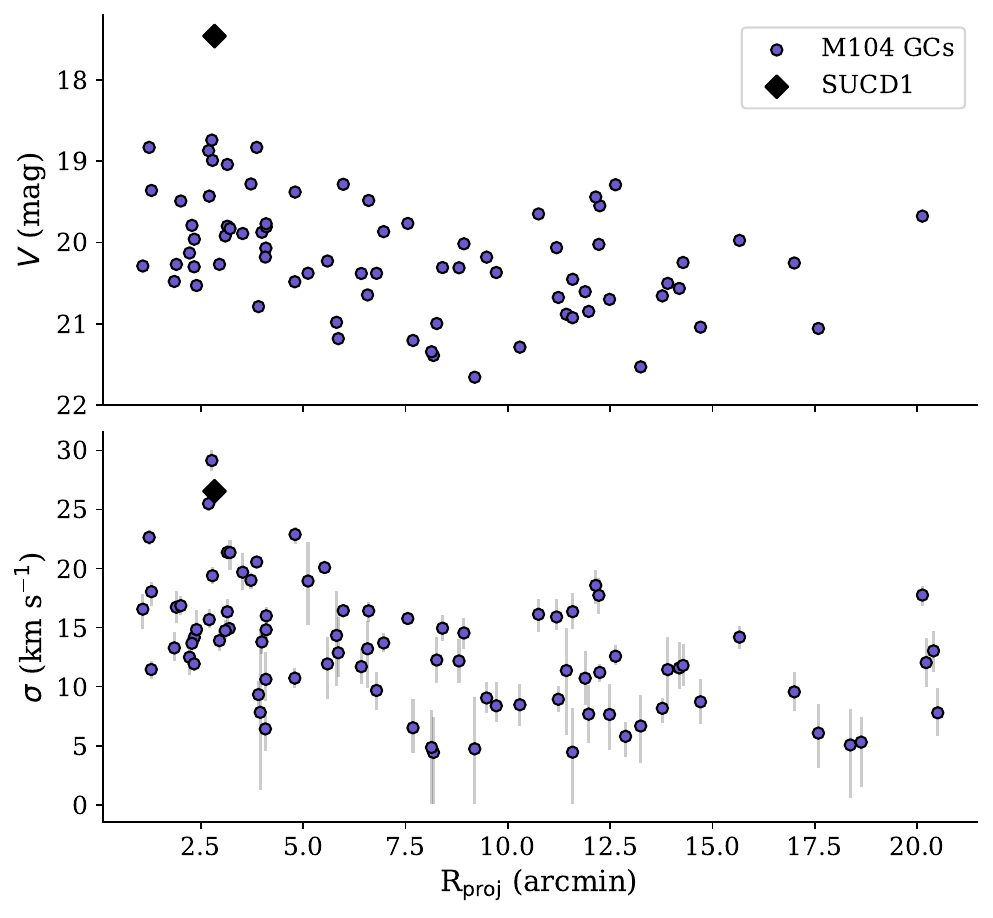}
    \caption{$V$-band magnitudes (top) and velocity dispersions (bottom) as a function of projected distance from the centre of M104. The diamond symbol refers to SUCD1.}
    \label{fig:rproject}
\end{figure}

Figure \ref{fig:rproject} shows the $V$-band magnitudes and velocity dispersions $\sigma$ of the M104 GCs as a function of projected distance from the centre, illustrating that some of the brightest GCs in the sample are also among the brightest. The close match in the $V$ versus $\sigma$ trend stems from the GCVD relation and we do not see any evidence for inflated velocity dispersions due to contamination from the galaxy light in the fiber apertures.

 \section{Table of dispersion measurements}
 \label{ap:ap}
We report the velocity dispersions, magnitudes, and where available sizes, mass-to-light ratios, and metallicities in Table \ref{tab:dispersions}. The identifiers refer to the catalogue presented in \cite{Fahrion2025b}.
\begin{table}[]
    \caption{Dispersion measurements.}
    \label{tab:dispersions}
    \centering
        \begin{tabular}{c c c c c c c c c c c}\hline\hline
        ID$^{1}$ & RA & DEC & $\sigma$ & $r_{\text{h}}$ & $V_{\text{Spitler+06}}$ &  $V_{\text{Harris+10}}$ & $V_{\text{Rhode\&Zepf04}}$ & $M_{\text{dyn}}/L_V$ & [M/H] & Metal ref$^{2}$. \\ 
     & (J2000) & (J2000) & (km s$^{-1}$) & (arcsec) & (mag) & (mag) & (mag) & ($M_\odot/L_\odot)$ & (dex) & \\ \hline
12 & 190.01304 & $-$11.66786 & $26.55^{0.35}_{-0.32}$ & 0.34 & ... & 17.46 & ... & 3.38$\pm$0.33 & $-$0.48$\pm$0.06  & FLAMES \\
13 & 190.03900 & $-$11.64231 & $25.49^{0.70}_{-0.61}$ & 0.12 & 18.91 & 18.87 & ... & 3.93$\pm$0.21 & $-$0.83$\pm$0.05  & FLAMES \\
14 & 190.04279 & $-$11.57125 & $10.62^{2.28}_{-1.80}$ & 0.07 & 20.11 & 20.07 & 20.14 & 1.16$\pm$0.47 & $-$1.76$\pm$0.11  & FLAMES \\
15 & 190.04479 & $-$11.78836 & $8.47^{1.75}_{-1.81}$ & ... & ... & ... & 21.29 & ... & $-$0.77$\pm$0.17  & FLAMES \\
16 & 190.04667 & $-$11.50436 & $6.52^{2.42}_{-2.17}$ & ... & ... & ... & 21.21 & ... & $-$1.14$\pm$0.24  & FLAMES \\
17 & 190.04687 & $-$11.84981 & $11.44^{2.78}_{-3.11}$ & ... & ... & ... & 20.50 & ... & $-$1.49$\pm$0.06  & FLAMES \\
18 & 190.05292 & $-$11.40906 & $6.66^{2.65}_{-3.10}$ & ... & ... & ... & 21.53 & ... & $-$0.43$\pm$0.23  & FLAMES \\
19 & 190.03271 & $-$11.38036 & $8.71^{1.97}_{-1.84}$ & ... & ... & ... & 21.04 & ... & $-$0.29$\pm$0.25  & FLAMES \\
20 & 190.05500 & $-$11.28775 & $13.03^{1.69}_{-1.78}$ & ... & ... & ... & ... & ... & $-$0.51$\pm$0.16  & FLAMES \\
21 & 190.07171 & $-$11.95222 & $12.04^{2.10}_{-2.10}$ & ... & ... & ... & ... & ... & $-$1.44$\pm$0.07  & FLAMES \\
23 & 190.09050 & $-$11.81750 & $5.79^{1.26}_{-1.72}$ & ... & ... & ... & ... & ... & $-$0.79$\pm$0.17  & FLAMES \\
24 & 190.09571 & $-$11.79056 & $4.45^{3.76}_{-4.37}$ & ... & ... & ... & 20.93 & ... & $-$1.79$\pm$0.17  & FLAMES \\
25 & 190.13242 & $-$11.93833 & $7.78^{2.07}_{-1.99}$ & ... & ... & ... & ... & ... & $-$1.75$\pm$0.08  & FLAMES \\
26 & 190.06908 & $-$11.55500 & $12.86^{3.02}_{-2.03}$ & ... & ... & ... & 21.18 & ... & $-$0.51$\pm$0.32  & FLAMES \\
27 & 190.03042 & $-$11.43528 & $11.36^{3.58}_{-5.45}$ & ... & ... & ... & 20.88 & ... & $-$2.15$\pm$0.15  & FLAMES \\
28 & 190.02520 & $-$11.59367 & $14.83^{1.67}_{-1.60}$ & 0.06 & 20.44 & 20.53 & 20.55 & 3.14$\pm$0.71 & $-$0.94$\pm$0.08  & OSIRIS \\
29 & 189.93025 & $-$11.81147 & $7.67^{2.49}_{-2.46}$ & ... & ... & ... & 20.85 & ... & $-$1.31$\pm$0.08  & FLAMES \\
30 & 189.95262 & $-$11.59462 & $21.37^{0.64}_{-0.60}$ & 0.10 & 19.09 & 19.04 & ... & 2.71$\pm$0.17 & $-$0.82$\pm$0.09  & OSIRIS \\
33 & 189.97587 & $-$11.57515 & $16.34^{1.06}_{-1.52}$ & 0.11 & 19.93 & 19.80 & 19.83 & 3.60$\pm$0.57 & $-$1.08$\pm$0.12  & OSIRIS \\
34 & 189.97846 & $-$11.68622 & $7.82^{6.88}_{-6.61}$ & ... & ... & ... & ... & ... & $-$1.14$\pm$0.41  & FLAMES \\
35 & 189.98367 & $-$11.71906 & $14.34^{3.75}_{-4.29}$ & ... & ... & ... & 20.98 & ... & $-$1.48$\pm$0.25  & FLAMES \\
36 & 190.02600 & $-$11.85794 & $11.58^{2.18}_{-1.76}$ & ... & ... & ... & 20.57 & ... & $-$1.69$\pm$0.14  & FLAMES \\
37 & 189.98450 & $-$11.53192 & $20.08^{0.39}_{-0.49}$ & ... & ... & ... & ... & ... & 0.00$\pm$0.08  & FLAMES \\
39 & 190.02096 & $-$11.48725 & $12.25^{1.98}_{-1.94}$ & ... & ... & ... & 21.00 & ... & 0.40$\pm$0.07  & FLAMES \\
40 & 190.02262 & $-$11.93269 & $5.30^{2.10}_{-3.76}$ & ... & ... & ... & ... & ... & $-$1.79$\pm$0.03  & FLAMES \\
41 & 190.02500 & $-$11.49014 & $4.85^{3.16}_{-4.81}$ & ... & ... & ... & 21.35 & ... & $-$1.49$\pm$0.17  & FLAMES \\
42 & 189.98825 & $-$11.92911 & $5.06^{3.03}_{-4.48}$ & ... & ... & ... & ... & ... & $-$1.33$\pm$0.09  & FLAMES \\
75 & 189.98917 & $-$11.64292 & $11.45^{0.74}_{-0.80}$ & ... & ... & ... & ... & ... & $-$1.85$\pm$0.20  & MUSE \\
91 & 189.99075 & $-$11.60258 & $18.04^{0.80}_{-1.19}$ & 0.07 & 19.39 & 19.36 & ... & 1.77$\pm$0.21 & $-$1.36$\pm$0.13  & MUSE \\
134 & 190.16358 & $-$11.75308 & $7.64^{2.57}_{-3.00}$ & ... & ... & ... & 20.70 & ... & $-$1.40$\pm$0.13  & FLAMES \\
135 & 190.16233 & $-$11.91731 & $17.75^{0.75}_{-0.97}$ & ... & ... & ... & 19.68 & ... & $-$1.07$\pm$0.13  & FLAMES \\
136 & 190.15450 & $-$11.76708 & $12.57^{0.92}_{-0.72}$ & ... & ... & ... & 19.29 & ... & $-$0.98$\pm$0.09  & FLAMES \\
137 & 190.15421 & $-$11.49092 & $18.57^{1.27}_{-1.56}$ & ... & ... & ... & 19.44 & ... & $-$1.06$\pm$0.15  & FLAMES \\
139 & 190.15133 & $-$11.68289 & $8.38^{2.05}_{-1.39}$ & ... & ... & ... & 20.37 & ... & $-$1.26$\pm$0.05  & FLAMES \\
143 & 190.14529 & $-$11.47961 & $17.73^{1.47}_{-1.57}$ & ... & ... & ... & 20.02 & ... & $-$1.06$\pm$0.14  & FLAMES \\
145 & 190.14083 & $-$11.75581 & $16.35^{1.58}_{-1.45}$ & ... & ... & ... & 20.45 & ... & $-$0.93$\pm$0.12  & FLAMES \\
154 & 190.11308 & $-$11.47386 & $8.93^{1.12}_{-1.10}$ & ... & ... & ... & 20.68 & ... & $-$0.71$\pm$0.16  & FLAMES \\
160 & 190.10167 & $-$11.89797 & $6.06^{2.47}_{-2.96}$ & ... & ... & ... & 21.06 & ... & $-$1.49$\pm$0.03  & FLAMES \\
168 & 190.07933 & $-$11.39892 & $11.79^{1.82}_{-1.78}$ & ... & ... & ... & 20.25 & ... & $-$0.92$\pm$0.12  & FLAMES \\
172 & 190.07371 & $-$11.67914 & $11.92^{2.26}_{-2.95}$ & ... & ... & ... & 20.23 & ... & $-$1.26$\pm$0.00  & FLAMES \\
181 & 190.06475 & $-$11.60589 & $6.42^{1.53}_{-1.91}$ & 0.08 & 20.21 & 20.18 & 20.19 & 0.61$\pm$0.33 & $-$1.12$\pm$0.13  & FLAMES \\
182 & 190.06408 & $-$11.62633 & $9.33^{1.15}_{-1.00}$ & 0.11 & 20.76 & 20.79 & 20.62 & 2.95$\pm$0.80 & $-$1.26$\pm$0.00  & FLAMES \\
184 & 190.06367 & $-$11.67028 & $22.88^{0.64}_{-0.79}$ & 0.08 & 19.37 & 19.38 & 19.33 & 3.58$\pm$0.23 & $-$0.91$\pm$0.05  & FLAMES \\
189 & 190.06083 & $-$11.53214 & $16.42^{0.75}_{-0.96}$ & ... & ... & ... & 19.48 & ... & $-$0.47$\pm$0.15  & FLAMES \\
192 & 190.05729 & $-$11.61739 & $19.69^{1.65}_{-1.53}$ & 0.04 & 19.80 & 19.89 & 19.91 & 2.08$\pm$0.34 & $-$0.69$\pm$0.14  & FLAMES \\
194 & 190.05654 & $-$11.65957 & $14.81^{1.24}_{-1.27}$ & 0.07 & 19.77 & 19.77 & 19.79 & 1.76$\pm$0.32 & $-$1.12$\pm$0.23  & OSIRIS \\
198 & 190.05179 & $-$11.42597 & $11.20^{0.72}_{-0.83}$ & ... & ... & ... & 19.55 & ... & $-$0.54$\pm$0.12  & FLAMES \\
210 & 190.04362 & $-$11.75575 & $14.95^{1.09}_{-1.04}$ & ... & ... & ... & 20.31 & ... & $-$0.43$\pm$0.17  & FLAMES \\
211 & 190.04333 & $-$11.61078 & $19.39^{0.70}_{-0.74}$ & 0.08 & 19.02 & 18.99 & 18.98 & 1.79$\pm$0.14 & $-$0.72$\pm$0.07  & FLAMES \\
215 & 190.04037 & $-$11.60633 & $15.68^{0.86}_{-0.72}$ & 0.09 & 19.57 & 19.43 & 19.45 & 1.86$\pm$0.20 & $-$1.28$\pm$0.09  & FLAMES \\
218 & 190.03883 & $-$11.58786 & $21.36^{1.02}_{-1.51}$ & 0.06 & 19.75 & 19.83 & 19.85 & 3.51$\pm$0.43 & $-$0.57$\pm$0.10  & FLAMES \\
221 & 190.03587 & $-$11.65861 & $14.74^{0.86}_{-0.72}$ & 0.09 & 19.94 & 19.92 & 19.85 & 2.66$\pm$0.32 & $-$0.76$\pm$0.08  & FLAMES \\
 \hline 

    \end{tabular}
\begin{tablenotes}
    \item $^{1}$ ID refers to the ID in the catalogue of \cite{Fahrion2025b}. ID 12 refers to SUCD1, magnitude of SUCD1 reported in \cite{Hau2009}. 
    \item $^{2}$ Refers to the instrument used for metallicity measurements as described in \cite{Fahrion2025b}.
\end{tablenotes}
\end{table}

\begin{table}[]
    \centering
    \begin{tabular}{c c c c c c c c c c c}\hline\hline
        ID & RA & DEC & $\sigma$ & $R_{\text{h}}$ & $V_{\text{Spitler+06}}$ &  $V_{\text{Harris+10}}$ & $V_{\text{Rhode\&Zepf04}}$ & $M_{\text{dyn}}/L_V$ & [M/H] & Metal ref.\\ 
     & (J2000) & (J2000) & (km s$^{-1}$) & (arcsec) & (mag) & (mag) & (mag) & ($M_\odot/L_\odot)$ & (dex) & \\ \hline
222 & 190.03537 & $-$11.72353 & $11.69^{1.46}_{-1.44}$ & ... & ... & ... & 20.38 & ... & $-$1.36$\pm$0.12  & FLAMES \\
226 & 190.03179 & $-$11.76808 & $14.54^{1.30}_{-1.31}$ & ... & ... & ... & 20.02 & ... & $-$0.90$\pm$0.11  & FLAMES \\
231 & 190.02949 & $-$11.60131 & $13.67^{0.56}_{-0.73}$ & 0.10 & 19.79 & 19.79 & 19.67 & 2.26$\pm$0.22 & 0.16$\pm$0.06  & OSIRIS \\
239 & 190.02342 & $-$11.54158 & $18.94^{3.34}_{-3.76}$ & ... & ... & ... & 20.38 & ... & $-$1.13$\pm$0.18  & FLAMES \\
240 & 190.02275 & $-$11.43819 & $15.90^{1.51}_{-1.10}$ & ... & ... & ... & 20.06 & ... & $-$0.50$\pm$0.21  & FLAMES \\
248 & 190.01504 & $-$11.48775 & $4.43^{3.03}_{-4.39}$ & ... & ... & ... & 21.39 & ... & $-$0.24$\pm$0.29  & FLAMES \\
249 & 190.01500 & $-$11.54497 & $10.72^{0.84}_{-0.86}$ & ... & ... & ... & 20.48 & ... & $-$0.10$\pm$0.14  & FLAMES \\
255 & 190.00900 & $-$11.50750 & $13.70^{0.82}_{-0.79}$ & ... & ... & ... & 19.87 & ... & $-$0.34$\pm$0.11  & FLAMES \\
264 & 190.00225 & $-$11.74897 & $15.77^{0.59}_{-0.51}$ & ... & ... & ... & 19.77 & ... & $-$0.21$\pm$0.12  & FLAMES \\
276 & 189.99562 & $-$11.58962 & $16.86^{0.83}_{-0.79}$ & 0.08 & 19.51 & 19.49 & 19.53 & 2.01$\pm$0.20 & $-$0.93$\pm$0.09  & OSIRIS \\
279 & 189.99167 & $-$11.57417 & $13.90^{0.89}_{-0.90}$ & 0.06 & 20.20 & 20.27 & 20.28 & 2.24$\pm$0.29 & $-$0.27$\pm$0.11  & FLAMES \\
281 & 189.98996 & $-$11.52369 & $16.43^{0.44}_{-0.43}$ & ... & ... & ... & 19.28 & ... & 0.40$\pm$0.01  & FLAMES \\
283 & 189.98700 & $-$11.33994 & $9.56^{1.69}_{-1.60}$ & ... & ... & ... & 20.25 & ... & $-$1.41$\pm$0.17  & FLAMES \\
288 & 189.98400 & $-$11.76928 & $12.17^{1.89}_{-1.91}$ & ... & ... & ... & 20.31 & ... & $-$1.49$\pm$0.04  & FLAMES \\
292 & 189.98221 & $-$11.65899 & $11.93^{0.83}_{-0.63}$ & 0.08 & 20.32 & 20.30 & 20.30 & 2.33$\pm$0.29 & $-$0.27$\pm$0.17  & OSIRIS \\
293 & 189.98083 & $-$11.55678 & $15.99^{0.73}_{-0.81}$ & ... & ... & ... & 19.81 & ... & $-$1.14$\pm$0.10  & FLAMES \\
302 & 189.97129 & $-$11.85125 & $8.16^{0.86}_{-1.24}$ & ... & ... & ... & 20.66 & ... & $-$1.11$\pm$0.15  & FLAMES \\
308 & 189.96717 & $-$11.63383 & $16.73^{1.33}_{-1.35}$ & 0.06 & 20.23 & 20.27 & 20.29 & 3.11$\pm$0.52 & $-$1.05$\pm$0.23  & FLAMES \\
320 & 189.96316 & $-$11.64266 & $14.17^{0.57}_{-0.63}$ & 0.11 & 19.92 & 19.96 & 19.70 & 3.23$\pm$0.32 & 0.24$\pm$0.05  & OSIRIS \\
339 & 189.94404 & $-$11.61414 & $14.92^{0.44}_{-0.46}$ & 0.10 & 19.84 & 19.82 & 19.68 & 2.81$\pm$0.32 & $-$0.05$\pm$0.10  & FLAMES \\
340 & 189.94371 & $-$11.58261 & $13.79^{1.26}_{-1.02}$ & ... & ... & ... & 19.88 & ... & $-$1.00$\pm$0.11  & FLAMES \\
346 & 189.93874 & $-$11.60014 & $19.00^{0.50}_{-0.72}$ & 0.13 & 19.33 & 19.28 & 19.13 & 3.67$\pm$0.26 & 0.16$\pm$0.04  & OSIRIS \\
355 & 189.92792 & $-$11.71342 & $9.67^{1.54}_{-1.63}$ & ... & ... & ... & 20.38 & ... & $-$1.49$\pm$0.06  & FLAMES \\
376 & 189.90033 & $-$11.56905 & $13.20^{2.35}_{-3.30}$ & ... & ... & ... & 20.65 & ... & $-$1.87$\pm$0.15  & OSIRIS \\
378 & 189.89896 & $-$11.79611 & $10.71^{2.35}_{-2.27}$ & ... & ... & ... & 20.61 & ... & $-$1.26$\pm$0.12  & FLAMES \\
381 & 189.89629 & $-$11.86453 & $14.19^{0.95}_{-1.01}$ & ... & ... & ... & 19.98 & ... & $-$1.10$\pm$0.11  & FLAMES \\
382 & 189.89621 & $-$11.50647 & $4.73^{4.42}_{-4.69}$ & ... & ... & ... & 21.66 & ... & 0.26$\pm$0.12  & FLAMES \\
387 & 189.88625 & $-$11.50875 & $9.04^{1.37}_{-1.24}$ & ... & ... & ... & 20.18 & ... & $-$0.69$\pm$0.13  & FLAMES \\
395 & 189.87221 & $-$11.49267 & $16.12^{1.27}_{-1.47}$ & ... & ... & ... & 19.65 & ... & $-$1.20$\pm$0.11  & FLAMES \\
412 & 190.04184 & $-$11.63909 & $29.15^{0.90}_{-0.92}$ & 0.08 & 18.79 & 18.74 & ... & 3.26$\pm$0.26 & $-$0.82$\pm$0.10  & OSIRIS \\
413 & 190.00021 & $-$11.60261 & $22.64^{0.63}_{-0.65}$ & 0.09 & 18.94 & 18.83 & ... & 2.20$\pm$0.30 & $-$0.88$\pm$0.10  & MUSE \\
414 & 190.06078 & $-$11.64111 & $20.55^{0.46}_{-0.36}$ & 0.11 & 18.98 & 18.83 & ... & 2.29$\pm$0.20 & $-$0.57$\pm$0.10  & OSIRIS \\
423 & 189.99700 & $-$11.64114 & $16.56^{1.30}_{-1.66}$ & 0.04 & 20.20 & 20.29 & ... & 1.97$\pm$0.40 & 0.22$\pm$0.08  & MUSE \\
424 & 189.96816 & $-$11.64635 & $12.49^{1.82}_{-1.49}$ & 0.09 & 20.20 & 20.13 & ... & 2.29$\pm$0.61 & $-$1.01$\pm$0.13  & OSIRIS \\
429 & 189.97517 & $-$11.60142 & $13.28^{1.38}_{-1.14}$ & 0.07 & 20.44 & 20.48 & ... & 2.82$\pm$0.73 & $-$0.36$\pm$0.23  & MUSE \\ \hline
    \end{tabular}

\end{table}
 \end{appendix}

\end{document}